\documentclass[prd,aps,floats,twocolumn,showpacs,nofootinbib]{revtex4}

\usepackage{amssymb}
\usepackage{amsmath}

\usepackage{graphicx}
\usepackage{graphics}
\usepackage{dcolumn}
\usepackage{color}
\usepackage{rotate}
\usepackage{fancyhdr} 

\def\be{\begin{equation}}
\def\ee{\end{equation}}
\def\bea{\begin{eqnarray}}
\def\eea{\end{eqnarray}}
\def\nn{\nonumber}

\def\veck{\mathbf{k}}

\def\VEV#1{\left\langle #1 \right\rangle}

\newcommand{\bfx}{\mathbf{x}}
\newcommand{\bfy}{\mathbf{y}}

\newcommand{\bfk}{\mathbf{k}}
\newcommand{\bfq}{\mathbf{q}}
\newcommand{\bgfk}{\mathbf{K}}
\newcommand{\bfn}{\mathbf{\hat{n}}}

\newcommand{\PhiOne}{\Phi^{(1)}}
\newcommand{\PsiOne}{\Psi^{(1)}}
\newcommand{\deltaOne}{\delta^{(1)}}
\newcommand{\vOne}{v^{(1)}}

\newcommand{\PhiTwo}{\Phi^{(2)}}
\newcommand{\PsiTwo}{\Psi^{(2)}}
\newcommand{\deltaTwo}{\delta^{(2)}}
\newcommand{\vTwo}{v^{(2)}}

\begin{document}

\title{Anisotropic imprint of long-wavelength tensor
     perturbations on cosmic structure}
\author{Liang Dai, Donghui Jeong, and Marc Kamionkowski}
\affiliation{Department of Physics \& Astronomy, Bloomberg
     Center, 3400 N.\ Charles Street, The Johns Hopkins University,
     Baltimore, MD 21218, USA}

\date{17 June 2013}

\begin{abstract}
 Inflationary models predict a correlation between primordial
 density perturbations (scalar metric perturbations) and
 gravitational waves (tensor metric perturbations) in the form of
 a scalar-scalar-tensor three-point correlation, or bispectrum in
 Fourier space. The squeezed limit of this bispectrum implies a
 quadrupolar asymmetry in the observed local power spectrum for
 matter and galaxies.  Here we show (like others before) that an
 infrared divergence in the amplitude of this power asymmetry
 predicted in single-field slow-roll models is canceled by
 projection effects when considering the observed power spectrum.
 We then further evaluate the nonzero, but finite, residual
 quadrupolar power asymmetry that remains after the divergences are canceled. 
 While the quadrupolar power asymmetry is small, it is conceptually important.
 Our calculation moreover clarifies how the
 predictions for this power asymmetry may change in models
 with different scalar-scalar-tensor bispectra, and shows
 that convincing detection of the quadrupolar power asymmetry would
 rule out the single-field slow-roll models of inflation.
\end{abstract}

\maketitle

\section{Introduction}
\label{sec:intro}

Three decades of increasingly precise measurements, culminating
most recently with those from the Planck satellite
\cite{Ade:2013uln}, have all shown consistency with the simplest
single-field slow-roll (SFSR) models of inflation
\cite{Guth:1982ec,Bardeen:1983qw,Hawking:1982cz,Linde:1981mu,Mukhanov:1981xt}.
Still, many questions about the new physics responsible for
inflation remain, and a number of further predictions of
inflation remain to be tested.  One of these predictions is a
stochastic background of gravitational waves, or tensor metric
perturbations
\cite{Abbott:1984fp,Rubakov:1982df,Fabbri:1983us,Starobinsky:1979ty}.
Efforts are now underway to detect these tensor modes in the
cosmic microwave background (CMB) polarization
\cite{Kamionkowski:1996zd,Seljak:1996gy}, and there are
prospects for direct detection of the background
\cite{Kawamura:2006up,Smith:2005mm,Smith:2008pf,Chongchitnan:2006pe}.

There may, however, also be an imprint of tensor modes in the
observed cosmic mass distribution.  One possible observable is
the distortion induced by gravitational lensing by tensor modes
to the  galaxy distribution \cite{Dodelson:2003bv,Schmidt:2012nw,Dai:2012bc},
the CMB
\cite{Cooray:2005hm,Li:2006si,Dodelson:2010qu,Book:2011na} or
the 21-cm background \cite{Pen:2003yv,Masui:2010cz,Book:2011dz}.
Another possibility---which
we focus upon here---is that long-wavelength tensor perturbations
may lead to a quadrupolar power asymmetry in the power spectra
for scalar perturbations.  The idea is simple:
tensor modes with wavelengths longer than the distance
over which observations are done give rise to a quadrupolar
distortion to the spacetime over the observed volume.  This
quadrupole may then (a) get imprinted somehow in the {\it
primordial} mass distribution and/or (b) induce a quadrupole in
the {\it observed} distribution through projection effects.
Such a quadrupole could then be sought, for example, in the
cosmic microwave background \cite{Pullen:2007tu,Hanson:2009gu}
or in galaxy surveys \cite{Ando:2008zza}.  In fact, null
searches for power asymmetries have already been carried out in
galaxy surveys \cite{Pullen:2010zy}.  A tentative detection in
the CMB \cite{Groeneboom:2008fz} was later disputed; current
measurements place upper limits
\cite{Bennett:2010jb,Ade:2013nlj} at the level of $\lesssim0.1$
on the amplitude of a power quadrupole.

One might think that the amplitude of the power asymmetry would
be large.  For example, in standard SFSR inflation, the
contribution to the square of the local tensor-perturbation amplitude
is equal across each logarithmic interval of tensor wavelength.
The root-mean-square (rms) of the local quadrupolar distortion then
scales with total number of $e$-folds of inflation, and this
number could conceivably be large.  An explicit calculation of
the scalar-scalar-tensor power quadrupole
\cite{Jeong:2012df,Dai:2013ikl}, based upon the SFSR
scalar-scalar-tensor bispectrum \cite{Maldacena:2002vr}, seems
to show such an infrared
divergence.  An upper limit on the power asymmetry
would then translate into an upper bound to the duration of
inflation.  Roughly speaking, the power quadrupole obtained this
way is $\sim N \gamma_{\rm rms}$,\footnote{We use $\gamma$
instead of $h$ for tensor modes.}
where $N$ is the number of
$e$-folds, and $\gamma_{\rm rms}\sim (\rho_{\rm infl}^{1/4}/m_{\rm Pl})^2
\lesssim 10^{-5}$ is the typical amplitude for a given
Fourier mode of the tensor field in terms of the energy density
$\rho_{\rm infl}$ during inflation and the Planck mass $m_{\rm
Pl}$.  A conservative current upper limit of $\lesssim 0.1$ for
the power quadrupole
\cite{Pullen:2010zy,Bennett:2010jb,Ade:2013nlj} then translates
to $N\lesssim 10^4$.

On second thought, there are several reasons to question
this result.  First of all, it seems strange that observables within
our horizon (which is contained within the last $N\lesssim 60$
$e$-folds of inflation) could be probing physics on scales many
orders of magnitude beyond the observable horizon.  This becomes
even clearer when we realize that as the tensor wavenumber $K\to0$,
a given Fourier mode of the tensor field approaches a constant
tensor perturbation $\gamma_{ij}$ to the metric.  However, the
coordinates in a metric with a constant tensor $\gamma_{ij}$ can
always be re-scaled to give a Friedman-Robertson-Walker metric.
In other words, observables only depend on (at least two) spacetime
derivatives, $\propto K^2$, of the tensor metric perturbation
$\gamma_{ij}$.  True, a perturbation of arbitrarily
long wavelength was presumably within the horizon at some
sufficiently early time during inflation.  But again, it seems
strange that observables within our horizon would depend on
asymptotically early times during inflation.  These arguments
thus suggest that the infrared divergence in the power
quadrupole is a gauge artifact.

Since Maldacena's paper~\cite{Maldacena:2002vr} on three-point
functions in inflation, a large literature (e.g.,
Refs.~\cite{Creminelli:2004yq,Creminelli:2011sq,Gerstenlauer:2011ti,Tanaka:2011aj,Lewis:2012tc,Flauger:2013hra})
has clarified that analogous divergences that arise
from the squeezed limit of the scalar-scalar-scalar bispectrum
are gauge artifacts.  Refs.~\cite{Maldacena:2002vr,Creminelli:2004yq}
point out that the only effect of the squeezed-limit bispectrum
in terms of the global synchronous-gauge coordinates used in
Ref.~\cite{Maldacena:2002vr} is a constant coordinate transform
into the local Fermi normal coordinates
(FNC). Ref.~\cite{Gerstenlauer:2011ti} splits perturbations into
small- and long-wavelength modes and absorbs the 
latter into the background.  Ref.~\cite{Tanaka:2011aj} defines
genuine gauge-invariant variables, whose bispectrum explicitly
vanishes in the squeezed limit. While there are some
differences (to be expanded upon below), many of those arguments
apply to the squeezed limit of the scalar-scalar-tensor
bispectrum.  As clarified most recently in
Ref.~\cite{Pajer:2013ana}, and also below, the divergence in the
primordial power quadrupole from the scalar-scalar-tensor
bispectrum is canceled precisely by projection effects induced
by the tensor perturbation at late times.

We show here, however, that there still remains in SFSR
inflation, after these divergences cancel, a nonzero but finite
quadrupolar asymmetry in the observed {\it local} power
spectrum. Here {\it local}  means that the power spectrum is
measured within a patch smaller than the wavelength  of the
tensor mode.\footnote{The quadrupolar asymmetry in the three-dimensional power spectrum
$P(\bfk)$ we focus on here should not be confused with the
quadrupolar component ($\ell=2$) of some two-dimensional angular
power spectrum $C_{\ell}$.}
While the switch, employed in previous work \cite{Pajer:2013ana}, 
to FNC coordinates right after inflation suffices to demonstrate the 
cancellation of divergences, it fails to account for the effects of spatial
and temporal variations of tensor perturbations
\cite{Giddings:2011zd}. Both FNC and
the peak-background split are only valid locally and hence
introduce ambiguities for finite Fourier wavelengths.  In this paper we
therefore work with global FRW coordinates but define observables in
a physical way.  For simplicity, we first derive rigorous results 
for a post-inflation Universe filled with non-relativistic matter. We then 
argue heuristically, but without a complete calculation, that
the power quadrupole induced by tensor modes of the smaller
wavelengths that enter the horizon during radiation domination
should be suppressed. The calculation presented
here thus applies to tensor modes outside the horizon and to modes
within the horizon today but that entered the horizon during matter
domination (i.e., with wavelengths
$\gtrsim70~\mathrm{Mpc}~h^{-1}$). This allows us to account for
the epoch of radiation domination that precedes matter
domination. Dark energy dominates only at very late times and
thus plays a marginal role on the power quadrupole. It does
affect the mapping between source redshift and comoving distance
though. With these insights, we are able to generalize our
results and make contact with our Universe.
Our work expands upon previous work \cite{Giddings:2011zd},
which also suggested a finite effect on the observed power
spectrum that grows logarithmically with time, by making precise
the prediction for {\it observable} quantities.

Our calculation shows that the observable power asymmetry
depends on spacetime derivatives of the tensor perturbation,
rather than just its amplitude.  As a result, the contribution from a given
superhorizon mode to the observable power asymmetry is
suppressed by a factor $K^2$ relative to the naive calculation.
Thus, local observables are not sensitive to
inflationary tensor modes with arbitrarily long wavelength. Our
conclusion is based on three considerations: (1) the squeezed
primordial scalar-scalar-tensor bispectrum satisfies the
single-field consistency relation 
[Eq.~(\ref{eq:sst-consistency-relation}) below]; 
(2) throughout the expansion
history after inflation, nonlinear mode couplings between
long-wavelength tensor perturbations and short-wavelength scalar
perturbations take effect; (3) the positions at which we
correlate matter/galaxy overdensities are specified in a {\it physical}
way, e.g. by the {\it observed} redshifts and the {\it observed}
angular position on the sky. 

We then find that the observable power quadrupole is induced at
any given time predominantly by tensor perturbations with
wavelengths comparable to the horizon at that time, as it should
be.  While the amplitude $\sim \gamma_{\rm rms}\lesssim
10^{-5}$ (rather than $N \gamma_{\rm rms}$) of the resulting
quadrupolar power asymmetry is too small to be observable today,
it is important conceptually to note that the power asymmetry
exists. We also note that in models where the self-consistency
relation is violated~\cite{BNSpaper},\footnote{Also for violations of the
scalar-scalar-scalar consistency relation, see, e.g.,
Ref.~\protect\cite{Flauger:2013hra,Namjoo:2012aa}}\ the power
quadrupole could conceivably be far larger. A null detection
can thus constrain such
alternative models. Conversely, detection of a power quadrupole
would rule out SFSR inflation.

We begin in Sec.~\ref{sec:notations} with a discussion of our
notations.  We then review in Sec.~\ref{sec:primordial} the
primordial scalar-scalar-tensor correlation.
Sec.~\ref{sec:post-inflation} considers the evolution of
scalar perturbations in the presence of a tensor perturbation.
Sec.~\ref{sec:galaxy-observed} then connects the
results of the previous Section to the {\it observed} galaxy
distribution, determined from redshifts and angular positions of
galaxies.  Sec.~\ref{sec:beyond-MD} comments on the
generalization of the calculation beyond the purely
matter-dominated case, and
Sec.~\ref{sec:quadrupole-galaxy-power-spectrum} then
evaluates numerically the amplitude of the quadrupolar power
asymmetry. Sec.~\ref{sec:concl} makes concluding remarks.
Many of the calculational details are presented in the
Appendixes.  Appendix~\ref{app:scalar-2pt-tensor} derives the
two-point correlation function for scalar perturbations in the
presence of a tensor mode.
Appendix~\ref{app:einstein-fluid-eqs} presents some details for
the derivation of the Einstein and fluid equations in the
Poisson gauge.  Appendix~\ref{app:nonlinear-corrections}
solves equations for the nonlinear evolution of density
perturbations in the presence of a tensor perturbation.
Appendix~\ref{app:lagrangian-coordinates} presents an
alternative derivation of the central result,
Eq.~(\ref{eq:delta-2nd-order}), by tracking with Lagrangian
coordinates a collection of freely-falling test particles.
Appendix~\ref{app:galaxy-observed} provides the mapping
between galaxy overdensities in general coordinates and observed coordinates,
and Appendix~\ref{app:galaxy-observed-power-spectrum}
derives the galaxy power spectrum in observed coordinates. Finally,
Appendix~\ref{app:galaxy-power-quadrupole-calc} collects useful
results for calculating the power quadrupole.

\section{Notation}
\label{sec:notations}

At the end of inflation, all perturbations of interest extend
well beyond the horizon. They are conveniently described in a
global comoving coordinate system, which allows for perturbative
calculations across all scales. We choose the Poisson 
gauge \cite{Bertschinger:1993xt}, in
which the perturbed Friedmann-Robertson-Walker (FRW) metric
reads,
\bea
\label{eq:perturbed-metric}
     ds^2 & = & - (1+2\Psi) dt^2 + 2 a(t) w_i dx^i dt \nn\\
      && + a^2(t) 
     (1+2\Phi) \left( \delta_{ij} + \gamma_{ij} \right) dx^i dx^j.
\eea
Here Latin $i,j,k\cdots=1,2,3$ indices are for three-dimensional
flat space, and they are raised and lowered by Kronecker deltas
$\delta^{ij}$ and $\delta_{ij}$. The two potentials $\Phi$ and
$\Psi$ encode scalar perturbations, while the traceless, divergence-free
$\gamma_{ij}$ encodes tensor perturbations. We
have chosen to factor out $(1+2\Phi)$ instead of writing
$\left(1+2\Phi\right)\delta_{ij}+\gamma_{ij}$. This ensures that
for tensor perturbations with infinite wavelength---i.e. a
constant $\gamma_{ij}$---the only effect is a trivial rescaling
of spatial coordinates $x^i \rightarrow x^i - \gamma^i_j
x^j/2$. The time-space $(0i)$ components of the
metric perturbations $w_i$ are divergence-free $\partial^i
w_i=0$. In the absence of primordial vector perturbations, $w_i$
only appears at second order~\cite{Boubekeur:2009uk}, and does
not affect the result for $\delta$ that we obtain here; we thus
neglect it hereafter.

The primordial values for metric perturbations are labeled with
a subscript p.
In later Sections, when we consider the projection effect, we
need to consider only the (linear) tensor metric perturbations,
\be
\label{eq:perturbed-metric-tensor-only}
     ds^2 = - dt^2 + a^2(t) \left( \delta_{ij} + \gamma_{ij} \right) dx^i dx^j.
\ee 

Lower-case $\bfk$'s
are reserved for wavevectors of scalar perturbations, and
uppercase $\bgfk$ for wavevectors of tensor
perturbations.  An overdot denotes the derivative with respect
to comoving time $t$, not to be confused with the conformal time
$\eta$. For conciseness, we suppress the time dependence
of variables whenever the suppression can induce no
ambiguity. It is then understood that the equations hold at any
given cosmic time.

Observed quantities (positions, perturbations, correlation
functions, etc.) will be labeled with a tilde, as opposed to
their counterparts computed in global FRW coordinates.
The precise meaning of ``observed'' will be elucidated in the Sections to
follow. 

The notation $\VEV{\cdots}_{\gamma}$ will be used for
correlations computed in the presence of a {\it given}
tensor-perturbation realization. This is to be distinguished
from correlations $\VEV{\cdots}_0$ computed for a cosmology {\it
without} tensor perturbations.

\section{Primordial scalar power spectrum with tensor perturbations}
\label{sec:primordial}

One way inflationary tensor perturbations can impact cosmic
structure formation is by distorting the primordial scalar
correlation function. Inflationary dynamics typically predict
correlations between a large-scale tensor mode with wavevector
$\bgfk$ and polarization $s=+,\times$, and two small-scale
scalar perturbations with wavevectors $\bfk_1$ and
$\bfk_2$. This is represented by a primordial
scalar-scalar-tensor bispectrum
$\VEV{\Phi_p(\bfk_1)\Phi_p(\bfk_2)\gamma_{p,s}(\bgfk)}$. In
many inflation scenarios, the bispectrum satisfies the {\it
consistency relation}~\cite{Maldacena:2002vr,Creminelli:2004yq}, 
\bea
\label{eq:sst-consistency-relation}
&& \VEV{\Phi_p(\bfk_1)\Phi_p(\bfk_2)\gamma_{p,s}(\bgfk)} \overset{K\rightarrow 0} {\longrightarrow} (2\pi)^3 \delta_D(\bfk_1 + \bfk_2 + \bgfk) \nn\\
&& \qquad \times \frac12 \frac{d\ln P_{\Phi}}{d\ln k} \epsilon^{ij}_s(\bgfk) \hat k_{1i} \hat k_{2j} P_{\gamma}(K) P_{\Phi}(k),
\eea
in the squeezed limit, up to model-specific corrections
suppressed by $\mathcal{O}(K^2/k^2)$. Here we define
$\bfk=(\bfk_2-\bfk_1)/2$.  The logarithmic derivative of the
power spectrum is $d\ln P_{\Phi}/d\ln k
= n_s - 4$, where $n_s$ is the scalar spectral index. The
bispectrum $\VEV{\Phi_p\Phi_p\gamma_p}$ implies that in a given
realization for $\gamma_{p,ij}$ of the tensor field, the
correlation between two $\Phi$ modes is~\cite{Jeong:2012df}
\bea
\label{eq:ss-power-spectrum-gamma}
     && \VEV{\Phi_p(\bfk_1)\Phi_p(\bfk_2)}_{\gamma} = (2\pi)^3
     \delta_D(\bfk_1+\bfk_2) P_{\Phi}(k) \nn\\
     && \qquad + \int \frac{d^3\bgfk}{(2\pi)^3} \sum_s (2\pi)^3
     \delta_D(\bfk_1+\bfk_2+\bgfk) \frac12 \frac{d\ln
     P_{\Phi}}{d\ln k} P_{\Phi}(k) \nn\\ 
     && \qquad \times \gamma^{*}_{p,s}(\bgfk)
     \epsilon^{ij}_s(\bgfk) \hat k_{1i} \hat k_{2j} +
     \mathcal{O}((K/k)^2),
\eea
where we have summed over the two tensor polarizations.

Fourier-transforming back to real space, we can derive a scalar
correlation function between two points $\bfx_1$ and $\bfx_2$,
with a separation $\bfx=\bfx_2-\bfx_1$ and a midpoint at
$\bfx_c=(\bfx_1+\bfx_2)/2$. A {\it local} scalar two-point
correlation, as a function of $\bfx_c$, is meaningful if the
correlation scale is small compared to the typical variation
scale of the tensor; i.e., $ K |\bfx| \ll 1$, or $K \ll k$ in
Fourier space.  In this regime (as derived in
Appendix~\ref{app:scalar-2pt-tensor}),
\bea
\label{eq:scalar-2pt-gamma}
     && \VEV{\Phi_p(\bfx_1) \Phi_p(\bfx_2)}_{\gamma} = \int
     \frac{d^3\bfk}{(2\pi)^3} e^{i\bfk\cdot\bfx} P_{\Phi}(k)
     \nn\\ 
     && \qquad\qquad \times \left[ 1 - \frac12 \frac{d\ln
     P_{\Phi}}{d\ln k} \gamma^{ij}_p \hat k_i \hat k_j +
     \mathcal{O}\left( \frac{\partial^2\gamma_p}{k^2} \right)
     \right], 
\eea
where $\gamma$ and its derivatives are evaluated at the midpoint
$\bfx_c$. Effectively, the primordial correlations between
large-scale tensor modes and small-scale scalar modes give rise
to an anisotropic primordial scalar power spectrum (cf. Eq.~(6)
in Ref.~\cite{Giddings:2011zd} and Eq.~(4.5) in
Ref.~\cite{Giddings:2010nc})
\bea
\label{eq:scalar-power-spectrum-gamma}
     \tilde P_{\Phi}(\bfk;\bfx_c) &=&  P_{\Phi}(k) \left[ 1 -
     \frac12 \frac{d\ln P_{\Phi}}{d\ln k} \gamma^{ij}_p (\bfx_c)
     \hat k_i \hat k_j \right. \nn \\
     & & \left. + \mathcal{O}\left(
     \frac{\partial^2\gamma_p}{k^2} \right) \right],
\eea
which applies to a local volume smaller than $\sim 1/K$ in the
vicinity of $\bfx_c$. 
The choice of the midpoint $\bfx_c$ has the advantage that the omitted 
corrections are at least second-order derivatives of $\gamma_{ij}$.

Naively, Eq.~(\ref{eq:scalar-power-spectrum-gamma}) suggests
that superhorizon tensor modes with arbitrarily long wavelengths
contribute to $\gamma^{ij}_p(\bfx_c)$, and hence induce a large
quadrupole in the scalar power spectrum. However, modes with
$K\to0$ lead to no observable effect. To see this, consider a constant
$\gamma^{ij}_p$, corresponding to a tensor mode of infinite
wavelength. One realizes that the same comoving separations
$|x^i|$ along different directions represent different {\it
physical} separations $|\tilde x^i| = |(\delta^i_j +
(\gamma_p)^i{}_{j})x^j|$ (can be defined by the proper distance,
or any other coordinate-independent measure of length), since
the tensor mode acts as an anisotropic background metric. The
anisotropy in the scalar two-point correlation function should
be measured by correlating pairs of points along different
directions but with the same {\it physical} separation. In terms
of physical positions, we can derive from
Eq.~(\ref{eq:scalar-2pt-gamma}),
\bea
\label{eq:statement-primordial-scalar-2pt}
     \VEV{\Phi_p\left(\tilde\bfx_1\right)
     \Phi_p\left(\tilde\bfx_2\right)}_{\gamma} =
     \VEV{\Phi_p\left(\tilde\bfx_1\right)
     \Phi_p\left(\tilde\bfx_2\right)}_{0}.
\eea
Note that in the absence of $\gamma_{ij}$ on the right-hand side,
$\tilde \bfx_{1,2}=\bfx_{1,2}$. This is to say that with
constant tensor perturbations we measure a physical correlation
function no different than what we would measure without. We
highlight the crucial role of the consistency relation that the
coefficient $(1/2)(d\ln P_{\Phi})/(d\ln k)$ in
Eq.~(\ref{eq:sst-consistency-relation}) ensures the validity of
Eq.~(\ref{eq:statement-primordial-scalar-2pt}).

Still, Eq.~(\ref{eq:statement-primordial-scalar-2pt}) receives
corrections, which represent genuine physical effects, of order
$(K/k)^2$, from the finiteness of the tensor-mode wavelength
\cite{Pajer:2013ana}.
Although our aim in this paper is to find the leading-order effect of the
long-wavelength tensor modes, here we neglect terms of order 
$(K/k)^2$, because these terms are much smaller compared to 
the $\mathcal{O}((K/aH)^2)$ correction that we will
discuss in Section~\ref{sec:galaxy-observed}.

Note that Eq.~(\ref{eq:scalar-power-spectrum-gamma}) is equivalent to 
the scalar-scalar-tensor bispectrum, Eq.~(\ref{eq:sst-consistency-relation}),
in its squeezed limit ($K\ll k$).
This equation operationally defines the role of long-wavelength tensor
modes: the long-wavelength tensor mode (wavenumber $K$)
centered at $\bfx_c$ modulates the \textit{local} scalar power spectrum 
(with wavenumber $k\gg K$) around the point.
Throughout this paper, we will express the effect of long-wavelength 
tensor modes in the same manner. That is, the specific situation that 
we are considering is 
in the presence of long-wavelength tensor modes
at a region centered around $\bfx_c$, and we are calculating the 
imprint of such tensor modes on the {\it observed} galaxy/matter
power spectrum.

\section{Post-inflationary evolution}
\label{sec:post-inflation}

Large-scale tensor perturbations can also affect structure
formation through nonlinear mode coupling during post-inflation
evolution.

To present our approach clearly, we first work
within a simplified cosmology, in which non-relativistic matter
(e.g., cold dark matter) dominates the energy density after
reheating. The matter component can be described as a fluid with
negligible pressure and anisotropic stress. Matter
perturbations, including the fractional density perturbation
$\delta$ and peculiar velocity $v_i$,
then grow from the primordial scalar perturbations. Still, we
will eventually consider the radiation-dominated epoch preceding
matter domination to better account for our Universe 
in Section~\ref{sec:beyond-MD}.

We start with the perturbed metric of
Eq.~(\ref{eq:perturbed-metric}) and write down the Einstein
equations $R^{\mu}{}_{\nu}-g^{\mu}{}_{\nu}R/2=8\pi
GT^{\mu}{}_{\nu}$ and fluid equations
$\nabla_{\mu}T^{\mu}{}_{\nu}=0$ for the metric and matter
perturbations, as detailed in
Appendix~\ref{app:einstein-fluid-eqs}. 

Our strategy is to treat scalar/matter perturbations (which are
small only in the linear regime) and tensor perturbations (which
are always small) as independent expansion parameters in the
perturbative expansion. In this spirit, we keep mixed
second-order terms of the order scalar/matter perturbations
multiplying the tensor perturbations, but ignore terms quadratic
in scalar/matter perturbation itself or in $\gamma_{ij}$
itself. The former pertains to the usual nonlinear structure
formation without tensor perturbations, which has been
extensively studied in the literature \cite{Bartolo:2010rw,Jeong:2010ag}. 
The latter, being practically negligible, is beyond the scope of this work.

We then decompose perturbations into a linear solution (labeled
${}^{(1)}$) plus a second-order correction (labeled ${}^{(2)}$),
e.g. $\Phi=\PhiOne+\PhiTwo$, and the same for $\Psi$, $\delta$,
$\gamma_{ij}$, and so on. For the peculiar velocity, we
decompose $v_i = \vOne_i + \vTwo_i + v_{R,i}$. Here, $\vOne_i$
and $\vTwo_i$ are the linear and second-order curl-free velocity
fields, respectively, and $v_{R,i}$ is the divergence-free
velocity field, which arises only in second order.

The linear solutions are
routinely solved in the standard linear cosmological
perturbation theory. The second-order corrections arise from
nonlinear mode-coupling between the tensor-perturbation modes
and the scalar/matter-perturbation modes. 

We are not interested in the nonlinear correction
$\gamma^{(2)}_{ij}$ for tensors, which describes
gravitational-wave emission from cosmic structures. Indeed, it
does not source second-order scalar/matter perturbations.
Therefore,
we cause no confusion by using $\gamma_{ij}$ in place of
$\gamma^{(1)}_{ij}$.

\subsection{Linear evolution}
\label{sec:linear-evolution}

The linear evolution of matter perturbations during the
matter-dominated epoch is easily solved.

At linear order, the peculiar velocity is curl-free.  In the
absence of anisotropic stress, the two scalar potentials
are related via $\PsiOne = -\PhiOne$, and $\PhiOne$ satisfies
the differential equation $\ddot \Phi^{(1)} + 4 H \dot
\Phi^{(1)} = 0$. Neglecting the decaying solution, the potential
is conserved, $\PhiOne=\Phi_p$, throughout
matter domination. Adiabatic initial conditions then lead to
linear growth for the matter perturbations (in Fourier space),
\bea
     && \deltaOne = 2 \mathcal{T}_{\delta}(k) \Phi_p , \quad
     \vOne = \frac{2}{3aH} \Phi_p,
\eea
where the linear-extrapolation factor for matter is
$\mathcal{T}_{\delta}(k)=1+k^2/(3a^2H^2)$.

At first order, $\gamma_{ij}$ evolves independently,
\be
\label{eq:tensor-evolve-linear}
     \ddot \gamma_{ij} + 3 H \dot \gamma_{ij} + a^{-2} k^2
     \gamma_{ij} = 0.
\ee
The solution (in Fourier space) is given by
$\gamma_{ij}=\mathcal{T}_{\gamma}(K) \gamma_{p,ij}$ with the
linear-extrapolation factor
$\mathcal{T}_{\gamma}(K)=3j_1(K\eta)/(K\eta)$ for tensor
modes. The tensor amplitude is conserved outside the horizon
and then oscillates and decays after horizon re-entry.

\subsection{Nonlinear tensor-scalar mode coupling}
\label{sec:mode-coupling}

In general, the nonlinear corrections for scalar/matter
perturbations satisfy the same linear, second-order differential
equations as the linear solutions do, but with inhomogeneous
source terms quadratic in linear solutions. 
They vanish at early times when perturbations are linear.

\begin{figure}[h]
\centering
\hspace{-0.5cm}
\includegraphics[scale=1]{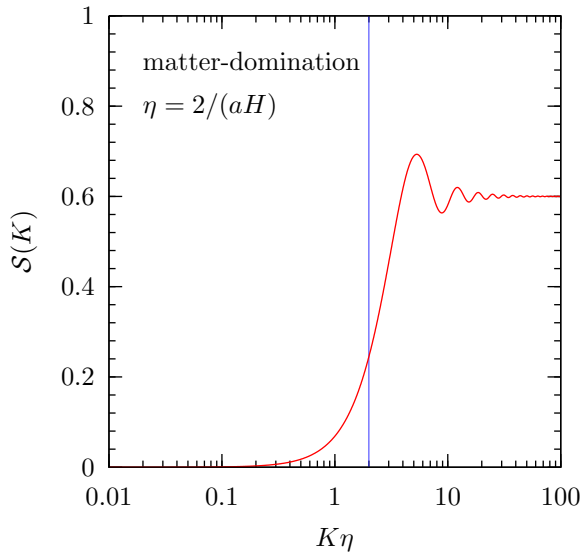}
\caption{The mode-coupling kernel $\mathcal{S}(K)$ is plotted assuming matter domination. It is only a function of the combination $K\eta$. The vertical line marks the comoving Hubble scale.}
\label{fig:Splot}
\end{figure}

In particular, the nonlinear correction to the potential
$\PhiTwo$, as derived in
Appendix~\ref{app:nonlinear-corrections}, can be solved from
\bea
\label{eq:2nd-Phi-equation}
     \ddot \Phi^{(2)}  + 4 H \dot \Phi^{(2)} =
     \frac{\partial^{-2}}{a^2} \left[ \left( \partial^2 \gamma_{ij} \right)
     \left( \partial^i \partial^j \PhiOne \right) \right] + H
     \delta\dot\Phi^{(2)}, \nn\\
\eea
where $\delta\PhiTwo\equiv\PhiTwo+\PsiTwo$ is given by
\bea
\label{eq:2nd-dPhi-result}
     \partial^2\delta\PhiTwo = 3 \partial^{-2} \left[ \left( \partial^2
     \gamma^{ij} \right) \left( \partial_i \partial_j \PhiOne\right) \right].
\eea
We now assume a scale hierarchy between the tensor perturbations
and the scalar/matter perturbations of interest---the typical
variation scale $\sim 1/K$ of tensors is much larger than the
scale $\sim1/k$ of scalar/matter perturbations; i.e., $K\ll
k$. In this regime, the variation of $\gamma_{ij}$ is
unimportant compared to that of $\PhiOne$, and we can apply the
inverse Laplacian only on the potential,
\bea 
     \partial^{-2} \left[ \left( \partial^2 \gamma^{ij} \right) \left( \partial_i
     \partial_j \PhiOne \right) \right] \approx \left( \partial^2
     \gamma^{ij}  \right) \left( \partial^{-2} \partial_i
     \partial_j \PhiOne \right).
\eea
Then according to Eq.~(\ref{eq:2nd-dPhi-result}), the second
term $H \delta\dot\Phi^{(2)}$ on the right-hand side of
Eq.~(\ref{eq:2nd-Phi-equation}) is $\sim H \dot\gamma K^2 k^{-2}
\PhiOne \sim K^2 \gamma \PhiOne (K/k)^2$, and hence is
$\mathcal{O}(K^2/k^2)$ smaller than the first term. Like in the
discussion of primordial scalar power spectrum in
Sec.\ref{sec:primordial}, we consistently ignore this term. The
solution to Eq.~(\ref{eq:2nd-Phi-equation}) can then be
obtained via a Green's function approach,
\bea
     \PhiTwo(t) & = & \int^t_0 dt' \frac{\partial^2
     \gamma^{ij}(t')}{a^2(t')} \left( \partial^{-2} \partial_i
     \partial_j \PhiOne(t') \right) \nn\\
     && \times \frac{2}{5H(t')} \left[ 1 -
     \left(\frac{a(t)H(t)}{a(t')H(t')} \right)^5 \right].
\eea
Then $\deltaTwo$, obtained from
Eq.~(\ref{eq:2nd-overdensity-algebraic}), can be combined with
$\deltaOne$ to give the full nonlinear matter overdensity
$\delta=\deltaOne+\deltaTwo$.  We work in Fourier space (consider
a single tensor mode with wavevector $\bgfk$ that constitutes a
realization for the tensor perturbation) and insert the various
results for linear solutions.  A final compact expression,
\bea
\label{eq:delta-2nd-order}
\delta & = & 2 \mathcal{T}_{\delta}(k) \left( 1 - \frac12 \frac{d\ln\mathcal{T}_{\delta}(k)}{d\ln k} \mathcal{T}_{\gamma}(K) \gamma^{ij}_p \hat k_i \hat k_j \right) \Phi_p  \nn\\
&&  - 2 \mathcal{T}_{\delta}(k) \gamma^{ij}_p \hat k_i \hat k_j  \Phi_p \mathcal{S}(K),
\eea
can be derived,
where $\gamma^{ij}_p$ is understood as the primordial tensor
perturbation evaluated on a comoving patch smaller than $1/K$,
over which small-scale matter-perturbation modes are
measured. In deriving this result, we have ignored
general-relativistic corrections that are suppressed by
either $(K/k)^2$ or $(aH/k)^2$ at observing time (see
Appendix~\ref{app:nonlinear-corrections}). As a check,
we present in Appendix~\ref{app:lagrangian-coordinates} an
alternative derivation of Eq.~(\ref{eq:delta-2nd-order}) using
Lagrangian coordinates for collisionless matter in the
subhorizon limit $k \gg aH$. The function $\mathcal{S}(K)$
(plotted in Fig.~\ref{fig:Splot}) is given by
\bea
     \mathcal{S}(K) = \int^{K\eta}_0 d(K\eta') \frac{K\eta'}{5}
     \mathcal{T}_{\gamma}(K\eta') \left[ 1 - \left(
     \frac{K\eta'}{K\eta} \right)^5 \right],
\label{eq:SK}
\eea 
which has asymptotic behaviors,
\bea
\label{eq:asympto-SofK}
     \mathcal{S}(K) & \simeq & \begin{cases}
     (1/14)(K\eta)^2,\qquad & K \ll 2/\eta, \\
     3/5, \qquad & K \gg 2/\eta.
\end{cases} 
\eea
Note that $K\eta=2$ corresponds exactly to the comoving horizon
scale $1/K=1/(aH)$. Therefore, anisotropic matter clustering
builds up only around the time of horizon re-entry of a given tensor mode.
Long before re-entry, the mode is superhorizon and its influence on
sub-horizon physics can be gauged away; long after re-entry, on
the other hand, the tensor amplitude redshifts away and can no
longer play a role.

Eq.~(\ref{eq:delta-2nd-order}) demonstrates that due to
large-scale tensor perturbations, matter undergoes anisotropic
clustering locally, and in Fourier space overdensity modes grow
with a quadrupolar dependence on the direction of the wavevector
$\bfk$. 

\subsection{Infinite wavelength tensor}
\label{sec:infinite-tensor}

We now examine the case of a constant tensor perturbation
$K\rightarrow 0$ to make the point that the nonlinear
corrections from mode coupling between tensor modes and
scalar/matter modes are indispensable to ensure a
gauge-invariant answer for an observable such as the correlation
function.

For constant $\gamma_{ij}$, we have $\mathcal{T}_{\gamma}(K)=1$
and $\gamma_{ij}=\gamma_{p,ij}$. For any observation at finite
$\eta$, $\mathcal{S}(K)$ vanishes as $K\rightarrow 0$. The
matter overdensity up to nonlinear order is simply
\bea
     \delta & = & 2 \mathcal{T}_{\delta}(k) \left( 1 - \frac12
     \frac{d\ln\mathcal{T}_{\delta}(k)}{d\ln k} \gamma^{ij}_p
     \hat k_i \hat k_j \right) \Phi_p.
\eea

The two-point correlation function for the matter overdensity in
comoving coordinates then reads
\bea
     && \VEV{\delta(\bfx_1)\delta(\bfx_2)}_{\gamma} \nn\\
     & = & \int d^3\bfk e^{i\bfk\cdot\bfx}  4
     \mathcal{T}^2_{\delta}(k) \left( 1 - \frac12
     \frac{d\ln\mathcal{T}^2_{\delta}(k)}{d\ln k} \gamma^{ij}_p
     \hat k_i \hat k_j \right) \tilde P_{\Phi} (\bfk) \nn\\ 
     & = & \int d^3\bfk e^{i\bfk\cdot\bfx}  4
     \mathcal{T}^2_{\delta}(k) \left( 1 - \frac12
     \frac{d\ln\mathcal{T}^2_{\delta}(k)}{d\ln k} \gamma^{ij}_p
     \hat k_i \hat k_j \right) \nn\\
     && \times \left( 1 - \frac12 \frac{d\ln P_{\Phi}}{d\ln k}
     \gamma^{ij}_p \hat k_i \hat k_j \right) P_{\Phi}(k) \nn\\ 
     & = & \int d^3\bfk e^{i\bfk\cdot\bfx} \left( 1 - \frac12
     \frac{d\ln P_{\delta}}{d\ln k} \gamma^{ij}_p \hat k_i \hat
     k_j \right) P_{\delta} (k),
\eea
where $\bfx=\bfx_2-\bfx_1$. We have kept terms up to linear
order in $\gamma_{ij}$ and have defined the isotropic matter
power spectrum $P_{\delta}(k)\equiv 4 \mathcal{T}^2_{\delta}(k)
P_{\Phi}(k)$, as would be found in the {\it absence} of tensor
perturbations. It clearly shows that tensor perturbations give
rise to an anisotropic matter power spectrum,
\bea
\label{eq:matter-power-spectrum-gamma}
     \tilde P_{\delta}(\bfk) & = & \left( 1 - \frac12 \frac{d\ln
     P_{\delta}}{d\ln k} \gamma^{ij}_p \hat k_i \hat k_j \right)
     P_{\delta} (k),
\eea
measured in comoving coordinates. This is the analog of
Eq.~(\ref{eq:scalar-power-spectrum-gamma}) in the
constant-$\gamma_{ij}$ limit. In particular, the nonlinear
correction $\deltaTwo$ contributes the
$d\ln\mathcal{T}_{\delta}/d\ln k$ term, which is needed to
combine with the primordial tilt to give the tilt  $d\ln
P_{\delta}/d\ln k$ of the matter power spectrum.

The same argument for the primordial scalar two-point correlation in
Sec.~\ref{sec:primordial} applies to the matter two-point
correlation---the {\it observed} correlation function should
be measured in physical length $\tilde x^i = (\delta^i_j +
(\gamma_p)^i{}_j)x^j$. Paralleling the derivation for
Eq.~(\ref{eq:statement-primordial-scalar-2pt}), we obtain from
Eq.~(\ref{eq:matter-power-spectrum-gamma}),
\bea
\label{eq:statement-matter-2pt}
     \VEV{\delta\left(\tilde\bfx_1\right)
     \delta\left(\tilde\bfx_2\right)}_{\gamma} =
     \VEV{\delta\left(\tilde\bfx_1\right)
     \delta\left(\tilde\bfx_2\right)}_{0}.
\eea
This is to say that with constant tensor perturbations we
measure a physical correlation function for matter overdensity
no different than what we would measure without.

For finite tensor wavelengths,
Eq.~(\ref{eq:statement-matter-2pt}) receives corrections with
derivatives of the tensor perturbation. 
Along with the contribution $S(K)$ from non-linear coupling,
the derivative terms will affect observables.
The derivative corrections, however, are of order $(K/k)^2$, which is smaller 
compared to $S(K)\propto (K\eta)^2$ on scales that we are interested in.
Therefore, we will neglect derivative corrections.

\section{Galaxy clustering in observed coordinates}
\label{sec:galaxy-observed}

In Sec.~\ref{sec:post-inflation} and
Sec.~\ref{sec:infinite-tensor}, we have argued {\it a priori}
that correlation functions measured in terms of some
``physical'' coordinates are more representative of actual
observations. In this Section, we justify the use of these
``physical'' coordinates by presenting an explicit construction
of them, following a coordinate-independent definition of
the correlation function. We also show that the correlation function
defined in that way is insensitive to infrared tensor modes.

\subsection{Projection effects from tensor perturbations}
\label{sec:galaxy-observed-projection}

Let us consider redshift surveys of galaxies as tracers of
matter. For simplicity, we assume a constant, linear galaxy bias
$b_g$, which relates the galaxy overdensity to matter overdensity
through $\delta_g=b_g \delta$.  
The linear bias provides a multiplicative factor in the power
spectrum, and it does not affect the resulting quadrupole.

In a redshift survey, the position of a galaxy is inferred from
its {\it apparent} position  $\hat n^i$ on the sky and its {\it
observed} redshift $z$, converted for a background cosmology
without metric perturbations. However, metric perturbations,
including the tensor perturbation, distort the photon
geodesic. As a result, the inferred position $\tilde\bfx$ and time
$\tilde t$, which we call observed
coordinates, differ from the original position  $\bfx$ and time
$t$ of the source, which are just the globally-defined
comoving position and comoving time,
\bea
     x^i = \tilde x^i + \Delta x^i, \qquad t = \tilde t + \Delta t,
\eea 
where $\Delta x^i$ and $\Delta t$ are first-order in metric
perturbations. We interpret observed coordinates $\tilde
x^i$ and $\tilde t$ as the ``physical'' position and time, since
they are the coordinates of the survey chart where we mark all
galaxies as we see them.

The projection effect from tensor perturbations, up to linear
order in $\gamma_{ij}$,\footnote{We do not
consider the contribution to the
projection effect from $\Phi$, as this is beyond the scope of our
discussion. It has been studied extensively in the literature
\cite{Yoo:2009au,Yoo:2010ni,Challinor:2011bk,Bonvin:2011bg,Jeong:2011as}
and can be taken into account separately if desired. In reality,
redshift-space distortions due to peculiar velocities are the 
major concern, and here we simply assume the effect can be modeled.}
can be calculated by tracing along a
null geodesic in the direction $\hat n^i$ at the origin (the
observer's location) back to redshift $z$, for the perturbed
metric Eq.~(\ref{eq:perturbed-metric-tensor-only}). The results are~\cite{Book:2010pf,Jeong:2012nu}
\bea
\label{eq:shift-time}
     \Delta t & = & \frac{1}{2H} \int^r_0 dr' \frac{\partial
     \gamma_{\parallel}}{\partial \eta}, \\
\label{eq:shift-parallel}
     \Delta x_{\parallel} & = & - \frac12 \int^r_0 dr'
     \gamma_{\parallel} - \frac{1}{2aH} \int^r_0 dr'
     \frac{\partial \gamma_{\parallel}}{\partial \eta}, \\
\label{eq:shift-perp}
     \Delta x^i_{\perp} & = & \frac r2 \left( \gamma^{ij}_{o}
     \hat n_j - \gamma_{o,\perp} \hat n^i \right) \nn\\
     && + \Pi^{ij} \int^r_0 dr' \left( \frac{r-r'}{2} \partial_j
     \gamma_{\perp} - \frac{r}{r'} \hat n^k \gamma_{jk} \right),
\eea
where we have decomposed $\Delta x^i = \hat n^i \Delta
x_{\parallel} + \Delta x^i_{\perp}$. The transverse part
satisfies $\Pi^i{}_j\Delta x^j_{\perp}=0$ with $\Pi^i{}_j
\equiv \delta^i_j - \hat n^i \hat n_j $. Also,
$\gamma_{\perp}\equiv \gamma_{ij}\hat n^i \hat n^j$, and
$r=|x^i|$ is the (zeroth-order) radial comoving distance to the
source galaxy. For all line-of-sight integrals, the integrand is
evaluated along the unperturbed geodesic $x^i = \hat n^i
(\eta_0-\eta)$, where $\eta_0$ is the conformal time
today. Moreover, variables labeled with a subscript o are
evaluated at the observer's location. 

For infinite tensor wavelength $K\rightarrow 0$,
$\gamma_{ij}=\gamma_{p,ij}$, $\Delta t$ vanishes, and $\Delta
x^i = - \gamma^{ij}_p x_j/2$.

\subsection{Galaxy overdensity in observed coordinates}
\label{sec:galaxy-observed-overdensity}

We can relate the galaxy overdensity in observed
coordinates to that in global coordinates using
conservation of the number of galaxies. To linear order in
$\gamma_{ij}$, we find (detailed in
Appendix~\ref{app:galaxy-observed}),
\bea
\label{eq:galaxy-overdensity-transform}
     \tilde \delta_g - \delta_g & = & \left( b_e H \Delta t +
     \partial_i \Delta x^i \right) + \left( \Delta_t \partial_t
     + \Delta x^i \partial_i \right) \delta_g \nn\\
     &&  +  \left( b_e H \Delta t + \partial_i \Delta x^i
     \right) \delta_g,
\eea
where the parameter $b_e \equiv ( d\ln a^3 n_g )/(d \ln a)$ can
be measured for a given galaxy sample. The first term exists even
without any intrinsic overdensity $\delta_g=0$, as it describes
the apparent galaxy overdensity due to 
the deflection of light emitted from galaxies.
We neglect this term here because a power quadrupole 
due to this term shows up only at quadratic order in the tensor amplitude.
The second term arises simply as a 
change of the galaxy density contrast due to the shift from 
the comoving coordinates of the galaxy to the ``observed'' coordinates.
The third term reflects the non-trivial distortion of the
volume element due to that shift.

Furthermore, the time derivative $\dot{\delta_g}=b_gH (d\ln
D/d\ln a)\delta \simeq b_gH\delta$ of the density contrast 
is smaller than the gradient $\partial_i\delta_g \simeq b_g k_i
\delta$ by $\mathcal{O}(H/k)$ of the density contrast,
since we observe scalar modes deep inside the horizon. We thus
simplify Eq.~(\ref{eq:galaxy-overdensity-transform}) as
\bea
\label{eq:galaxy-overdensity-observed}
     \tilde \delta_g - \delta_g & = & \Delta x^i \partial_i
     \delta_g  +  \left( b_e H \Delta t + \partial_i \Delta x^i
     \right) \delta_g \nn\\
     & = & b_g \left[ \Delta x^i \partial_i \delta  +  \left(
     b_e H \Delta t + \partial_i \Delta x^i \right) \delta
     \right]. 
\eea
Having derived this formula, we hereafter remove the tilde from
coordinates since all quantities now refer directly to the corresponding
observables.

\subsection{Local power spectrum in observed coordinates}
\label{sec:galaxy-observed-power-spectrum}

Using Eq.~(\ref{eq:galaxy-overdensity-observed}), 
we calculate the galaxy power spectrum
(detailed in Appendix~\ref{app:galaxy-observed-power-spectrum}) 
in the vicinity of $\bfx_c$, in the presence of a single long-wavelength 
tensor mode with wavevector $\bgfk$. We find that as a result of
the long-wavelength tensor mode, 
locally an anisotropic galaxy power spectrum measured in
observed coordinates arises,
\bea
\label{eq:local-galaxy-power-spectrum-result}
     \tilde P_g(\bfk;\bfx_c) & = & b^2_g \left[ \tilde
     P_{\delta}(\bfk;\bfx_c) - \left(\partial_j \Delta
     x_i\right) \hat k^i \hat k^j \frac{d\ln P_{\delta}(k)}{d\ln
     k} P_{\delta}(k) \right.\nn\\
     && \left. + \left( 2 b_e H \Delta t + \partial_i \Delta x^i
     \right) P_{\delta}(k)  \right],
\eea
up to linear order in $\gamma_{ij}$. Note that from one volume
to another, the value of $\gamma_{ij}$ varies, and hence the local
power spectrum depends on the central position $\bfx_c$
of the local volume. From Eq.~(\ref{eq:delta-2nd-order}), the
local matter power spectrum $\tilde P_{\delta}$ reads
\bea
     && \tilde P_{\delta}(\bfk;\bfx_c) = 4
     \mathcal{T}^2_{\delta}(k) \left[ \tilde
     P_{\Phi}(\bfk;\bfx_c) \right.\nn\\
     &&\left.- \left( \frac12 \frac{d\ln
     \mathcal{T}^2_{\delta}(k)}{d\ln k} \mathcal{T}_{\gamma}(K)
     +2 \mathcal{S}(K) \right) \gamma^{ij}_p \hat k_i \hat k_j
     P_{\Phi}(k) \right].\nonumber \\
\label{eq:tilde_Pdelta_kxc}
\eea
We then insert Eq.~(\ref{eq:scalar-power-spectrum-gamma}) for
the primordial scalar power spectrum, and ignore the
$\mathcal{O}((K/k)^2)$ term as we have ignored terms of the same
order in Sec.~\ref{sec:mode-coupling}. We obtain an
expression for the effect of a single tensor mode with
wavevector $\bgfk$,
\bea
\label{eq:anisotropic-galaxy-power-spectrum}
     && \tilde P_g(\bfk;\bfx_c) = P_{g}(k) \left[ 1 + \left( 2
     b_e H \Delta t + \partial_i \Delta x^i \right)
     \right.\nn\\
     && \left. - \frac{d\ln P_{\Phi}(k)}{d\ln k} \hat k^i \hat
     k^j \left( \partial_j \Delta x_i + \frac12 \gamma_{p,ij}
     \right) \right. \nn\\
     && \left. - \frac{d\ln \mathcal{T}^2_{\delta}(k)}{d\ln k}
     \hat k^i \hat k^j \left( \partial_j \Delta x_i + \frac12
     \gamma_{ij} \right) - 2 \mathcal{S}(K) \gamma^{ij}_p \hat
     k_i \hat k_j \right], \nn\\
\eea
for the local galaxy power spectrum, 
where $P_{g}(k) \equiv 4 b^2_g \mathcal{T}^2_{\delta}(k)
P_{\Phi}(k)$ is the isotropic galaxy power spectrum that would
be observed in the absence of tensor perturbations. Correction
terms in the square brackets, except for the first term,
introduce a quadrupolar dependence on the direction of $\bfk$. 
We refer the reader to our final results,
Eq.~(\ref{eq:quadrupole-tensor}) and
Eq.~(\ref{eq:quadrupole-tensor-FNC}),  for the galaxy power
quadrupole, at which point we provide physical interpretations
for each term from the perspective of both the global comoving
frame and the locally FRW-like frame.

Now we check the superhorizon limit $K\rightarrow 0$ of
Eq.~(\ref{eq:anisotropic-galaxy-power-spectrum}).  From
Eq.~(\ref{eq:shift-time})-(\ref{eq:shift-perp}), we have in that
limit $\Delta t \rightarrow 0$ and $\partial_i \Delta x^i
\rightarrow 0$ so the second term vanishes.  Furthermore,
because $\partial_j \Delta x_i \rightarrow - \gamma_{p,ij}/2$
and because $\gamma_{ij}=\mathcal{T}_{\gamma}
\gamma_{p,ij}\rightarrow \gamma_{p,ij}$, the third and the
fourth term are identically zero in that limit. Finally, from
Eq.~(\ref{eq:asympto-SofK}) we have $\mathcal{S}(K) \propto
K^2$, so the last term vanishes as well. We thus conclude that
$K=0$ superhorizon tensor modes from inflation, which are
constant within our Hubble volume, have no observable effect on
galaxy clustering, as long as the consistency relation
Eq.~(\ref{eq:sst-consistency-relation}) holds. In fact, for
small $K$ not even terms linear in $K$ survive, and the leading
contribution goes as $\sim K^2$.

We highlight that the quadrupole in the power spectrum is
coherently induced for all small-scale matter modes (i.e. it is
$k$-independent), because in typical models $P_{\Phi}(k)$ and
$\mathcal{T}_{\delta}(k)$ take power-law forms to good
approximation. 

\section{Beyond matter domination}
\label{sec:beyond-MD}

The anisotropic galaxy power spectrum, 
Eq.~(\ref{eq:anisotropic-galaxy-power-spectrum}), 
has been derived analytically under the assumption that the Universe 
is matter dominated right after inflation.
In reality, however, an epoch of radiation-domination precedes the
matter-dominated era. 

Nevertheless, one can still generalize
Eq.~(\ref{eq:anisotropic-galaxy-power-spectrum}) to take
radiation domination into account. 
To do so, we go through a second-order analysis paralleling that
in Sec.~\ref{sec:post-inflation}, but now assume radiation to be
the major component of the energy-stress tensor. During
radiation domination, the sub-horizon modes of the potential
$\Phi$ oscillate and decay due to radiation pressure. Induced
by tensor perturbations, the anisotropic part of the local power
spectrum for these modes also oscillates over time and scale. 
Consequently, the local potential power spectrum does not
develop a quadrupole that is coherent for all scalar-mode wavenumber
$k$'s, contrary to the case of matter-domination. Furthermore,
the sub-dominant dark-matter component only grows
logarithmically in response to the potential, and does not
develop an anisotropic power spectrum coherent over many scales
either. Therefore, we simply assume that {\it no} anisotropy in
the scalar/matter power spectrum builds up during radiation
domination. In principle, more detailed numerical calculations
can quantitatively account for the radiation-matter transition,
but we will leave this for future work.

With this physical picture, we include the effects of radiation
domination by nominally taking the amplitude of the
scalar/matter perturbations at the radiation-to-matter transition as
the ``primordial'' amplitude. This makes no difference for
scalar/matter modes that re-enter the horizon after 
matter-radiation equality,
but reduces the ``initial'' amplitude for modes that re-enter
earlier, as 
their linear growth is retarded during radiation domination.
Effectively, the linear-extrapolation factor
$\mathcal{T}_{\delta}(k)$ in
Eq.~(\ref{eq:anisotropic-galaxy-power-spectrum}) must have a
turnover,
\bea
     \mathcal{T}_{\delta}(k) \simeq \begin{cases}
     k^2/(3a^2H^2),\qquad & k<k_{\mathrm{eq}}, \\
     k_{\mathrm{eq}}^2/(3a^2H^2),\qquad & k>k_{\mathrm{eq}}.
\end{cases}
\eea
where $k_{\mathrm{eq}}$ corresponds to the Hubble scale at
matter-radiation equality.

As has been pointed out, a given tensor mode only induces
anisotropy in the matter power spectrum around the time of
horizon re-entry through nonlinear mode-coupling, and no
anisotropy builds up until matter domination. Hence, the
$\mathcal{S}(K)$ term in
Eq.~(\ref{eq:anisotropic-galaxy-power-spectrum}) is cut off on
the small-scale end at $K \sim k_{\mathrm{eq}}$. On the
large-scale end, the effective cut-off is the horizon scale at
source redshift.

Not until very recently in the cosmic history does dark energy
dominate the Universe. For simplicity, we completely
ignore its effects on the evolution of both scalar/matter and
tensor perturbations, but only account for its geometrical
effect when the source redshift is converted to the comoving
distance.

\section{Quadrupole of the galaxy power spectrum}
\label{sec:quadrupole-galaxy-power-spectrum}

We now quantify the level of anisotropy in the local galaxy
power spectrum. The anisotropic distortion can be described by
five quadrupole moments, 
\bea
     \mathcal{Q}_{2m}(\bfx_c) \equiv \frac{\int d^2\hat \bfk
     \tilde P_g(\bfk;\bfx_c) Y^*_{(2m)}(\hat\bfk)}{\int d^2\hat
     \bfk \tilde P_g(\bfk;\bfx_c) Y^*_{(00)}(\hat\bfk)},
\eea
for $m=\pm2,\pm1,0$, with $Y_{(\ell m)}(\hat\bfk)$ defined with
respect to some chosen coordinate axes. Using
Eq.~(\ref{eq:anisotropic-galaxy-power-spectrum}), we find
\bea
     \mathcal{Q}_{2m} (\bfx_c) = \int d^2\hat\bfk \mathcal{Q}_{ij}(\bfx_c)
     \left(\hat k^i \hat k^j - \frac13 \delta^{ij} \right)
     Y^*_{(2m)}(\hat\bfk),
\eea
where the contribution to the symmetric, traceless quadrupole
tensor $\mathcal{Q}_{ij}$ from a single tensor mode of wavenumber $K$
reads
\bea
\label{eq:quadrupole-tensor}
     && \mathcal{Q}_{ij} = - \frac12 \frac{d\ln
     P_{\Phi}}{d\ln k} \gamma_{p,ij} - \left( \frac{\mathcal{T}_{\gamma}(K)}{2} \frac{d\ln
     \mathcal{T}^2_{\delta}}{d\ln k} + 2
     \mathcal{S}(K) \right) \gamma_{p,ij} \nn\\
     && - \left( \frac{d\ln P_{\Phi}}{d\ln k} + \frac{d\ln
     \mathcal{T}^2_{\delta}}{d\ln k} \right) \left(
     \partial_{(i} \Delta x_{j)} - \frac13 \delta_{ij}
     \partial\cdot\Delta x \right).
\eea
This provides a perspective in comoving coordinates: the first
term gives the naive prediction from the squeezed limit of the
SFSR scalar-scalar-tensor bispectrum; the second term arises
from nonlinear
mode coupling; and the last term accounts for the projection 
into the observed coordinates. Although we have suppressed the dependence on source
position $\bfx_c$, remember that $\partial_i \Delta x_j$ is evaluated at location
 $\bfx_c$, and the value of the quadrupole differs from patch to patch. 
Another equivalent form (see
Appendix~\ref{app:lagrangian-coordinates}) is
\bea
\label{eq:quadrupole-tensor-FNC}
     && \mathcal{Q}_{ij} = 
     - \frac12 \frac{d\ln P_{\delta}}{d\ln k} 
       \left( 1 - \mathcal{T}_{\gamma}\right) \gamma_{p,ij} 
            + 2 \mathcal{S}_N(K) \gamma_{p,ij}  \nn\\
     && - \frac{d\ln P_{\delta}}{d\ln k} 
     \left(
		\frac12 \gamma_{ij} + 
     \partial_{(i} \Delta x_{j)} - \frac13 \delta_{ij}
     \partial\cdot\Delta x \right),
\eea
where only the matter power spectrum $P_{\delta}$ explicitly
enters. Here the function $\mathcal{S}_N(K)$ is given explicitly
in Eq.~(\ref{eq:kernel-SN-def}). This form provides a physical
interpretation in a {\it locally} FRW-like
frame~\cite{Pajer:2013ana}: the first line is due to the
residual tidal forces in that frame created by long-wavelength
tensor perturbations; the second line, proportional to
$(1/2)\gamma_{ij} + \partial_{(i} \Delta x_{j)} - (1/3)
\partial\cdot\Delta x$ at source location and time, describes
the gauge-invariant projection effect \cite{Schmidt:2012nw}.

We predict not the definite values of
the quadrupole moments, but only their root-mean-square,
\bea
     \overline{\mathcal{Q}^2} & \equiv & \VEV{ \sum_{m=-2}^2
     |\mathcal{Q}_{2m}|^2 },
\eea
which is orientation-independent. It can be shown that
$\overline{\mathcal{Q}^2}=(8\pi/15)
\VEV{\mathcal{Q}_{ij}\mathcal{Q}^{ij*}}$.

We assume that the primordial $\gamma_{p,ij}$ is a realization
of a Gaussian random field, which is statistically homogeneous
and isotropic, with a power spectrum parameterized by
\bea
\label{eq:tensor-power-spectrum-primordial}
     \VEV{\gamma_{p,s}(\bgfk) \gamma_{p,s'}(\bgfk')} = (2\pi)^3
     \delta_D(\bgfk+\bgfk') P_{\gamma}(K) \delta_{ss'}.
\eea
Then the rms of the quadrupole only depends on the distance from
the observer to the source, or equivalently the source redshift $z$,
but not on the angular direction in the sky. Using the
total-angular-momentum formalism~\cite{Dai:2012bc}, we find
\bea
\label{eq:expression-quadrupole-rms}
     \overline{\mathcal{Q}^2}(z) = \frac{4}{15\pi}
     \int^{K_{\mathrm{max}}} _0 K^2dK P_{\gamma}(K) \sum_{J=2}^{\infty} (2J+1) \mathcal{Q}^2_J(K,z). \nn\\
\eea
The expressions for the coefficient function
$\mathcal{Q}^2_J(K,z)$ are provided in
Appendix~\ref{app:galaxy-power-quadrupole-calc}. In practice,
the summation over the total-angular-momentum quantum number $J$
is truncated for $J \gtrsim Kr$ with $r$ the comoving distance
to the source. The wavenumber integral is subject to a cutoff
$K_{\mathrm{max}}$, which should satisfy $K_{\mathrm{max}}\ll
k$. For measuring the galaxy power quadrupole on scales
$k>k_{\mathrm{eq}}$, we have $d\ln P_{\Phi}/d\ln k=n_s-4$ and
$d\ln\mathcal{T}_{\delta}/d\ln k=0$. In that case, we can choose
$K_{\mathrm{max}} \sim k_{\mathrm{eq}}$. 

For numerical evaluation, we take the flat $\Lambda$CDM
concordance cosmology with the WMAP+BAO+$H_0$ best-fit
cosmological parameters of Ref.~\cite{Komatsu:2010fb}. A
scale-free primordial power spectrum
$P_{\gamma}(K)=2\pi^2\Delta^2_{\gamma}/K^3$ is expected from
inflation. We discuss the treatment of the radiation- and
dark-energy--dominated epochs in Sec.~\ref{sec:beyond-MD}.

\begin{figure}[h]
\centering
\hspace{-0.5cm}
\includegraphics[scale=1]{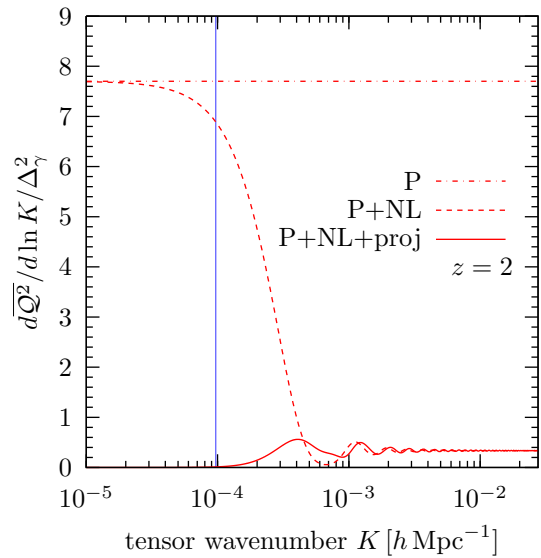}
\caption{The contribution to $\overline{\mathcal{Q}^2}$ from per
     logarithmic interval of $K$ for source redshift $z=2$,
     normalized to $\Delta^2_{\gamma}$. The dash-dotted line (P)
     is the prediction from SFSR scalar-scalar-tensor bispectrum alone.
     (first term of Eq.~(\protect\ref{eq:quadrupole-tensor})). The
     dashed line (P+NL) includes nonlinear mode couplings (second
     term of Eq.~(\protect\ref{eq:quadrupole-tensor})). The solid line
     (P+NL+proj) is the full result with the projection effects
     (last term of Eq.~(\protect\ref{eq:quadrupole-tensor})). The
     vertical line marks the horizon scale at present time.} 
\label{fig:Kplot}
\end{figure}

\begin{figure}[h]
\centering
\hspace{-0.5cm}
\includegraphics[scale=1]{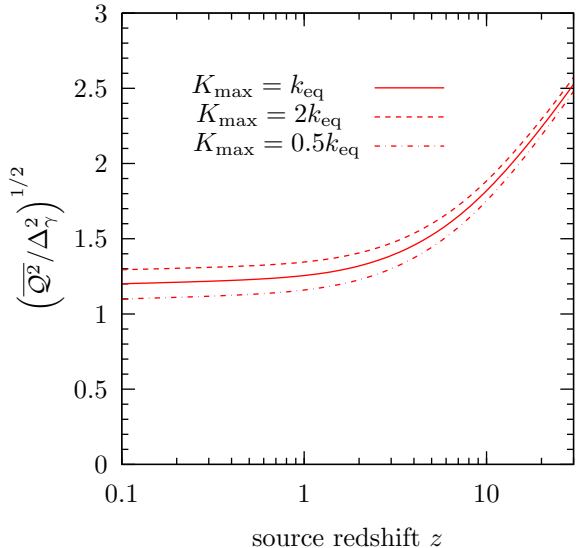}
\caption{The variance of the galaxy power quadrupole as a
     function of the source redshift $z$, normalized to
     $(\Delta^2_{\gamma})^{1/2}$. We compare results for
     different choices of the small-scale cutoff
     $K_{\mathrm{max}}$ around the scale $k_{\mathrm{eq}}$.}
\label{fig:zplot}
\end{figure}

Fig.~\ref{fig:Kplot} gives an example (at $z=2$) of the
contribution to the rms of the quadrupole per logarithmic
interval of the tensor wavenumber $K$. On subhorizon scales, the
nonlinear mode-coupling effects partially cancel with the
prediction from the primordial bispectrum alone. 
The constant limit for large $K$ reflects that each tensor mode
$K$ induces the same {\it cumulative} effect (as evident from
the $K\eta\rightarrow \infty$ limit in Fig.~\ref{fig:Splot}) on
small-scale galaxy quadrupole from its horizon re-entry to  
damp-out, as long as $K \ll k$. On the other hand,
the projection effects kick in on superhorizon scales to cancel
the primordial contribution, and hence ensure infrared-safety
(without projection, the quadrupole will be proportional to
superhorizon $e$-folds). Therefore, the residual quadrupole from
the full result is dramatically smaller. Fig.~\ref{fig:zplot}
shows the variance of the quadrupole as a function of the galaxy
redshift. We truncate the tensor wavenumber at
$K=k_{\mathrm{eq}}$, but also show that a factor-of-two
variation of that choice only modifies the result
marginally. Over a wide range of redshifts ($0.1\lesssim
z\lesssim3$) accessible to galaxy surveys the variance is $\sim
1.2 (\Delta^2_{\gamma})^{1/2}$. 

\section{Conclusion}
\label{sec:concl}

In this paper, we have derived the quadrupolar asymmetry
imprinted on the galaxy (or other tracers of the matter) power
spectrum due to long-wavelength tensor perturbations in
single-field slow-roll (SFSR) inflation.
We have considered the case of a scale hierarchy between the matter
inhomogeneities of interest and the tensor perturbations, so the
power quadrupole can be measured from many local volumes, across
each of which the tensor perturbation marginally varies.

The observed quadrupole is the sum of three contributions: (1) A
primordial quadrupole is imprinted at the end of inflation from
a primordial curvature-curvature-tensor bispectrum, satisfying
the squeezed-limit consistency relation. (2) An extra quadrupole
develops during post-inflationary evolution, because matter
clusters in the anisotropic background due to the long-wavelength
tensor modes. This is a cosmic-scale analogy of the intrinsic
alignment of galaxies due to tensor perturbations~\cite{Schmidt:2012nw}.
(3) A third contribution arises from projection
into observed coordinates, which are defined by the
observed redshift and the apparent angular position of the source. 
Unlike the amplitude of the primordial quadrupole due to
tensor modes, which is sensitive to superhorizon $e$-folds, the
sum of the three contributions to {\it observable} effects is
insensitive to tensor modes that are superhorizon today. We
therefore conclude that the squeezed-limit consistency relations
for primordial bispectra guarantee that superhorizon
perturbations generated by SFSR inflation have no observable
consequences on subhorizon physics. The residual effects are
due to modes that are subhorizon or are undergoing horizon re-entry
today. We have thus quantified in terms of precisely defined
observables the magnitude of the observable quadrupole in
the matter power spectrum suggested in Ref.~\cite{Giddings:2011zd}.

With the current constraint on the inflationary tensor amplitude,
the power quadrupole on scales $k<k_{\mathrm{eq}}$ is $\lesssim
10^{-5}$ in single-field slow-roll inflation. On smaller scales
$k>k_{\mathrm{eq}}$ the amount is expected to differ, but not
significantly. Given the plenty of ``real-world'' complications
such as redshift-space distortion and nonlinear evolution of
matter perturbations, the imprints from tensors are beyond the
reach of current large-scale-structure surveys. 
Nominally, this signal is an order of magnitude larger than the 
tensor-mode signals induced in the angular power spectra of galaxy
clustering~\cite{Jeong:2012nu} and shear~\cite{Schmidt:2012nw}
through projection effects. Still, with future galaxy and 21-cm surveys, 
the quadrupolar power asymmetry might eventually provide a competitive 
probe of tensor modes from large-scale-structure surveys.  
Moreover, detection of the power
asymmetry does not necessarily require a full-sky survey, as the
overall signal-to-noise depends on the three-dimensional survey
volume.  Therefore, high-redshift accessibility can compensate
for a moderate angular coverage.

If an inflation model violates the consistency relation
Eq.~(\ref{eq:sst-consistency-relation}), e.g. by having a
different scaling law with respect to $K$ (preferably redder
for phenomenological interest), then the infrared-safety of the
observed power quadrupole will not hold. Such a scenario may predict
a large quadrupole due to the abundance of superhorizon tensor
modes, and hence will be subject to stringent constraints from
observations.
Conversely, detection of a large quadrupolar asymmetry in the
galaxy power spectrum, beyond the level predicted in this work,
would rule out single-field slow-roll inflation.

\begin{acknowledgments}
We thank Fabian Schmidt for useful discussions and comments on
an earlier draft.  L.D.\ is grateful for the support of the
William Gardner Fellowship.  This work was supported by DoE
SC-0008108 and NASA NNX12AE86G. 
\end{acknowledgments}

\begin{widetext}

\appendix

\section{Primordial scalar two-point correlation in the presence of tensor}
\label{app:scalar-2pt-tensor}

The primordial two-point correlation function for the scalar
potential $\Phi$ is the inverse Fourier-transform of
Eq.~(\ref{eq:ss-power-spectrum-gamma}),
\bea
\label{eq:app1-1}
   \VEV{\Phi_p(\bfx_1)\Phi_p(\bfx_2)}_{\gamma} & = & \int
   \frac{d^3\bfk_1}{(2\pi)^3}\, \frac{d^3\bfk_2}{(2\pi)^3}
   e^{i(\bfk_1\cdot\bfx_1+\bfk_2\cdot\bfx_2)}
   \VEV{\Phi_p(\bfk_1)\Phi_p(\bfk_2)}_{\gamma} \nn\\
   & = & \int \frac{d^3\bfk}{(2\pi)^3} e^{i\bfk\cdot\bfx}
   P_{\Phi}(k) + \int \frac{d^3\bfk_1}{(2\pi)^3}\,
   \frac{d^3\bfk_2}{(2\pi)^3}
   e^{i(\bfk_1\cdot\bfx_1+\bfk_2\cdot\bfx_2)} \int
   \frac{d^3\bgfk}{(2\pi)^3} \sum_s (2\pi)^3
   \delta_D(\bfk_1+\bfk_2+\bgfk) P_{\Phi}(k) \nn\\
   && \times \left\{ \frac12 \frac{d\ln P_{\Phi}}{d\ln k}
   \gamma^*_{p,s}(\bgfk) \epsilon^{ij}_s(\bgfk) \hat k_{1i} \hat
   k_{2j} + \mathcal{O} ((K/k)^2) \right\},
\eea
where we define $\bfq=\bfk_1+\bfk_2$ and
$\bfk=(\bfk_2-\bfk_1)/2$. For the second equality, the first
term is the homogeneous and isotropic correlation in the absence
of $\gamma_{ij}$. For the second term, we plug in $(2\pi)^3
\delta_D(\bfk+\bfk'+\bgfk)=\int d^3\bfy
\exp[i(\bfk+\bfk'+\bgfk)\cdot\bfy]$. Changing to
$\bfx_c=(\bfx_1+\bfx_2)/2$ and $\bfx=\bfx_2-\bfx_1$, we have
\begin{align}
\label{eq:app1-2}
     & \int \frac{d^3\bfk_1}{(2\pi)^3}\,
     \frac{d^3\bfk_2}{(2\pi)^3}
     e^{i(\bfk_1\cdot\bfx_1+\bfk_2\cdot\bfx_2)} \int
     \frac{d^3\bgfk}{(2\pi)^3} \sum_s (2\pi)^3
     \delta_D(\bfk_1+\bfk_2+\bgfk) P_{\Phi}(k) \left\{ \frac12
     \frac{d\ln P_{\Phi}}{d\ln k} \gamma^*_{p,s}(\bgfk)
     \epsilon^{ij}_s(\bgfk) \hat k_{1i} \hat k_{2j} +
     \mathcal{O} ((K/k)^2) \right\} \nn\\
     = & \int \frac{d^3\bfk_1}{(2\pi)^3}\,
     \frac{d^3\bfk_2}{(2\pi)^3}
     e^{i(\bfk_1\cdot\bfx_1+\bfk_2\cdot\bfx_2)} \int
     \frac{d^3\bgfk}{(2\pi)^3} \sum_s \int d^3\bfy
     e^{i(\bfk_1+\bfk_2+\bgfk)\cdot\bfy} P_{\Phi}(k) \left\{
     \frac12 \frac{d\ln P_{\Phi}}{d\ln k} \gamma^*_{p,s}(\bgfk)
     \epsilon^{ij}_s(\bgfk) \hat k_{1i} \hat k_{2j} +
     \mathcal{O} ((K/k)^2) \right\} \nn\\
     = & \int \frac{d^3\bfk}{(2\pi)^3}\,
     \frac{d^3\bfq}{(2\pi)^3}
     e^{i(\bfq\cdot\bfx_c+\bfk\cdot\bfx)} \int
     \frac{d^3\bgfk}{(2\pi)^3} \sum_s \int d^3\bfy e^{i(\bfq
     +\bgfk)\cdot\bfy} P_{\Phi}(k) \left\{ \frac12 \frac{d\ln
     P_{\Phi}}{d\ln k} \gamma^*_{p,s}(\bgfk)
     \epsilon^{ij}_s(\bgfk) \hat k_{1i} \hat k_{2j} +
     \mathcal{O} ((K/k)^2) \right\}.
\end{align}
The $\bfy$ integral fixes $\bfq=-\bgfk$, and $d\ln P_{\Phi}/d\ln
k = n_s-4$ is just a number. Moreover, under the assumption $K
\ll k$ we have the Taylor expansion,
\bea
\label{eq:app1-3}
     \epsilon^{ij}_s(\bgfk) \left. \hat k_{1i} \hat k_{2j}
     \right|_{\bfq=-\bgfk} & = & - \epsilon^{ij}_s(\bgfk) \hat
     k_i \hat k_j + \mathcal{O}((K/k)^2), 
\eea
because $\epsilon^{ij}_s(\bgfk) K_i =0$. We then integrate out
$\bfy$ and $\bfq$ to obtain
\bea
\label{eq:app1-4}
     && \int \frac{d^3\bfk}{(2\pi)^3}\, \frac{d^3\bfq}{(2\pi)^3}
     e^{i(\bfq\cdot\bfx_c+\bfk\cdot\bfx)} \int
     \frac{d^3\bgfk}{(2\pi)^3} \sum_s \int d^3\bfy \exp[i(\bfq
     +\bgfk)\cdot\bfy] P_{\Phi}(k) \left\{ - \frac12 \frac{d\ln
     P_{\Phi}}{d\ln k} \gamma^*_{p,s}(\bgfk)
     \epsilon^{ij}_s(\bgfk) \hat k_i \hat k_j + \mathcal{O}
     ((K/k)^2) \right\} \nn\\
     = && \int \frac{d^3\bfk}{(2\pi)^3} P_{\Phi}(k) \int
     \frac{d^3\bgfk}{(2\pi)^3} \sum_s
     e^{i(-\bgfk\cdot\bfx_c+\bfk\cdot\bfx)} \left\{ - \frac12
     \frac{d\ln P_{\Phi}}{d\ln k} \gamma^*_{p,s}(\bgfk)
     \epsilon^{ij}_s(\bgfk) \hat k_i \hat k_j + \mathcal{O}
     ((K/k)^2) \right\} \nn\\
     = && \int \frac{d^3\bfk}{(2\pi)^3} e^{i \bfk\cdot\bfx }
     P_{\Phi}(k) \left\{ - \frac12 \frac{d\ln P_{\Phi}}{d\ln k}
     \gamma^{ij}_p(\bfx_c) \hat k_i \hat k_j + \mathcal{O}
     (\partial^2\gamma/k^2) \right\}.
\eea
In the last line, we have used the definition of the Fourier
decomposition for $\gamma_{ij}$. Together with the
homogeneous/isotropic term in Eq.~(\ref{eq:app1-1}), this gives
Eq.~(\ref{eq:scalar-2pt-gamma}). 

\section{Einstein and fluid equations}
\label{app:einstein-fluid-eqs}

Here we present some key results in deriving the Einstein
equations and the fluid equations in the Poisson gauge. We treat
scalar/matter perturbations as independent small parameters from
tensor perturbations, and at second-order we only keep their
cross terms.  We consistently assume that there is no primordial
vector perturbations; thus, $w_i$ and $v_{R,i}$ are
$\mathcal{O}(\Phi\gamma)$.

The Levi-Civita connection coefficients are given by
\bea
\Gamma^0_{00}  & = & \dot\Psi, \qquad
\Gamma^0_{0i}  = \Gamma^0_{i0}   =  \partial_i\Psi -a H w_i, \qquad
\Gamma^i_{00}  = h^{ij} \partial_j\Psi + \frac{1}{a} \left( \dot w^i + H w^i \right), \nn\\
\Gamma^i_{j0}  = \Gamma^i_{0j} & = & H\delta^i_j + \frac12 \dot \gamma^i_j + \dot\Phi \delta^i_j + \frac{1}{2a} \left( \partial_j w^i - \partial^i w_j \right), \nn\\
\Gamma^0_{ij}  & = &   H h_{ij} +\frac{a^2}{2} \dot\gamma_{ij} + \left( 2H(\Phi-\Psi) +\dot\Phi \right) h_{ij} + a^2 (\Phi-\Psi)\dot\gamma_{ij} - \frac{a}{2} \left( \partial_i w_j + \partial_j w_i \right), \nn\\
\Gamma^k_{ij}  & = & \frac12 \left( \partial_i \gamma^k_j + \partial_j \gamma^k_i - \partial^k\gamma_{ij} \right) + \left[ \delta^k_j \partial_i \Phi + \delta^k_i \partial_j \Phi -h_{ij} h^{kl} \partial_l \Phi \right] + a H w^k \delta_{ij}, \nn
\eea
where we define $h_{ij}\equiv a^2(\delta_{ij}+\gamma_{ij})$ and $h^{ij}\equiv a^{-2}(\delta^{ij}-\gamma^{ij})$. The Ricci tensor is given by
\bea
R^0{}_0 & = & 3 \left( \dot H + H^2 \right) + \left[ 3 \ddot\Phi + 6 H \dot\Phi - 3 H \dot\Psi - 6 \left( \dot H + H^2 \right) \Psi - h^{ij}\partial_i\partial_j\Psi \right], \nn\\
R^0{}_i & = & \left( 2\partial_i\dot\Phi -  2 H \partial_i\Psi \right) - \frac12 \dot\gamma^j_i \partial_j \left( 3\Phi - \Psi \right) + 2 a H \left( \dot w_i - 2 H w_i \right) + \frac{1}{2a} \partial^2 w_i, \nn\\
R^i{}_0 & = & - h^{ij} \left( 2\partial_j\dot\Phi -  2 H \partial_j\Psi \right) + \frac12 a^{-2} \dot\gamma^{ij}\partial_j \left( 3\Phi - \Psi \right) - \frac{2H}{a} \left( \dot w^i - \frac{\dot H}{H} w^i - 2 H w^i \right) - \frac{a}{2} \partial^2 w^i, \nn\\
R^i{}_j & = & \,\left( \dot H + 3 H^2 \right) \delta^i_j  + \frac12 \left[ \ddot \gamma^i_j + 3 H \dot \gamma^i_j - a^{-2} \partial^2 \gamma^i_j \right]  \nn\\
         && + \left[ \ddot\Phi + H\left( 6\dot\Phi - \dot\Psi \right) - 2 \left( \dot H + 3 H^2 \right) \Psi \right] \delta^i_j - h^{ik}\partial_k\partial_j \left( \Phi + \Psi \right) - \delta^i_j h^{kl}\partial_k\partial_l\Phi   \nn\\
         &&  - \Psi \ddot\gamma^i_j - \dot\gamma^i_j \left[ H \left( 2\Phi + \Psi \right) + \frac12 \left( \dot\Phi + \dot\Psi \right)\right] + \Phi \partial^2 \gamma^i_j  + \frac12 \left( \partial^i\gamma^k_j + \partial_j\gamma^{ik} - \partial^k\gamma^i_j \right) \left( \partial_k \Phi + \partial_k \Psi \right) \nn\\
&& + \frac{1}{2a} \left( \partial^i \dot w_j + \partial_j \dot w^i \right).
\eea

For a Universe dominated by non-relativistic matter with energy
density $\rho_m$, matter perturbations are parameterized by
overdensity $\delta$, and peculiar velocity $v_i$ (as measured
by observers at fixed comoving position). Pressure and
anisotropic stress can be neglected. The matter energy-stress
tensor is given by
\bea
T^0{}_0  =  - \rho_m \left( 1+\delta \right), \quad
T^0{}_i  =  a \rho_m \left( v_i - w_i \right), \quad
T^i{}_0  =  - a \rho_m h^{ij} \left( v_j - w_j \right), \quad
T^i{}_j  =  0.
\eea

The Einstein equations $R^{\mu}{}_{\nu}-g^{\mu}{}_{\nu}R/2=8\pi GT^{\mu}{}_{\nu}$ for perturbations can be then obtained,
\bea
\label{eq:Einstein-time-time-perturb}
&& - h^{ij}\partial_i\partial_j \Phi + 3 H \dot\Phi - 3 H^2 \Psi = 4\pi G \rho_m \delta, \\
\label{eq:Einstein-time-space-perturb}
&& \left( 2\partial_i\dot\Phi -  2 H \partial_i\Psi \right) - \frac12 \dot\gamma^j_i \partial_j \left( 3\Phi - \Psi \right) + 2aH \left( \dot w_i - 2 H w_i \right) + \frac{1}{2a} \partial^2 w_i = 8 \pi G a \rho_m \left( v_i - w_i \right), \\
\label{eq:Einstein-space-space-perturb}
&&           \frac12 \left[ \ddot \gamma^i_j + 3 H \dot \gamma^i_j - a^{-2} \partial^2 \gamma^i_j \right]    + \left[ -2\ddot\Phi + H \left( -6\dot\Phi + 2\dot\Psi \right) + h^{kl}\partial_k\partial_l \left( \Phi + \Psi \right) \right] \delta^i_j - h^{ik}\partial_k\partial_j\left( \Phi + \Psi \right) - \Psi \ddot\gamma^i_j  + \Phi \partial^2 \gamma^i_j    \nn\\
&&            - \dot\gamma^i_j \left[ H \left( 2\Phi + \Psi \right) + \frac12 \left( \dot\Phi + \dot\Psi \right)\right]  + \frac12 \left( \partial^i\gamma^k_j + \partial_j\gamma^{ik} - \partial^k\gamma^i_j \right) \left( \partial_k \Phi + \partial_k \Psi \right) + \frac{1}{2a} \left( \partial^i \dot w_j + \partial_j \dot w^i \right) = 0.
\eea

The fluid equations for perturbations can be derived from $\nabla_{\mu}T^{\mu}{}_{\nu}=0$ and by applying background evolution equations. They read
\bea
\label{eq:continuity-perturb}
\dot\delta + a h^{ij} \partial_i v_j + 3 \dot\Phi & = & 0, \\
\label{eq:Euler-perturb}
\dot v_i + H v_i - \left( \dot w_i + 2 H w_i \right)  +  \frac1a \partial_i\Psi & = & 0.
\eea

Radiation needs to be added as an independent component when
discussing the epoch of radiation-domination. We assume radiation is
not coupled to matter (as for cold dark matter), but in the meantime has
negligible higher moments (as it will if tightly coupled to a
small amount of baryons). Then it can be described as a fluid
with energy density $\rho_r$ (with perturbation $\delta_r$),
pressure $p_r=\rho_r/3$, and curl-free bulk velocity
$v_{ri}=\partial_i v_r$. Together with non-relativistic matter,
the total energy-stress tensor reads,
\bea
T^0{}_0 & = & - \rho_r \left( 1+\delta_r \right) - \rho_m \left( 1+\delta \right), \quad
T^0{}_i  =  \frac43 a \rho_r \left( v_{ri} - w_i \right) + a \rho_m \left( v_i - w_i \right), \nn\\
T^i{}_0 & = & - \frac43 a \rho_r h^{ij} \left( v_{rj} - w_j \right) - a \rho_m h^{ij} \left( v_j - w_j \right), \quad
T^i{}_j  =  \frac13 \rho_r (1+\delta_r) \delta^i_j.
\eea
With these equations, the corresponding Einstein equations and
fluid equations can be derived straightforwardly.

\section{Equations for nonlinear corrections}
\label{app:nonlinear-corrections}

A complete set of differential equations for the nonlinear
corrections $\PhiTwo$, $\PsiTwo$, $\deltaTwo$ and $\vTwo_i$ can
be obtained by extracting the second-order part of the Einstein
equations,
Eqs.~(\ref{eq:Einstein-time-time-perturb}) and
(\ref{eq:Einstein-time-space-perturb}) (take the
divergence)
and Eq.~(\ref{eq:Einstein-space-space-perturb}) (take the
trace), and the fluid equations,
Eqs.~(\ref{eq:continuity-perturb}) and
(\ref{eq:Euler-perturb}) (take the curl-free part). They can be
simplified by applying the
background evolution $2\dot H + 3H^2=0$ and the evolution
equations for linear solutions.
We collect them here,
\bea
\label{eq:2nd-Einstein-time-time}
-\frac{\partial^2}{a^2} \PhiTwo + 3 H \dot\Phi^{(2)} - 3 H^2 \PsiTwo - 4 \pi G \rho_m \deltaTwo & = & - \gamma_{ij} \frac{\partial^i \partial^j}{a^2} \PhiOne, \\
\label{eq:2nd-Einstein-time-space}
\frac1a \partial^2 \dot\Phi^{(2)} - H \frac1a \partial^2 \PsiTwo - 4\pi G \rho_m \partial\cdot\vTwo & = & \dot\gamma^{ij} \frac1a \partial_i \partial_j \PhiOne, \\
\label{eq:2nd-Einstein-space-space}
-2\ddot\Phi^{(2)} + H \left( -6 \dot\Phi^{(2)} + 2\dot\Psi^{(2)} \right) + \frac23 a^{-2} \partial^2 \left( \PhiTwo + \PsiTwo \right) & = & 0, \\
\label{eq:2nd-continuity}
\dot\delta^{(2)} + \frac1a \partial\cdot \vTwo + 3 \dot \Phi^{(2)} & = & \gamma^{ij} \frac1a \partial_i \vOne_j, \\
\label{eq:2nd-Euler}
\dot v^{(2)}_i + H \vTwo_i + \frac1a \partial_i \PsiTwo & = & 0.
\eea
We note that $w_i$ and $v_{R,i}$ drop out of these equations;
as far as the matter overdensity is concerned, they can be
ignored.

At nonlinear order, the difference between two scalar potentials
$\delta\PhiTwo\equiv \PhiTwo+\PsiTwo$ does not
vanish. Eliminating $\PsiTwo$ in terms of this symbol, we derive
from Eq.~(\ref{eq:2nd-Einstein-space-space}),
\bea
\label{eq:2nd-Phi-trace}
\ddot\Phi^{(2)}  + 4 H \dot \Phi^{(2)} & = & \frac{1}{3a^2} \partial^2 \delta\PhiTwo + H \delta\dot\Phi^{(2)}.
\eea
Alternatively, we rewrite Eq.~(\ref{eq:2nd-Einstein-time-space}) and Eq.~(\ref{eq:2nd-Euler}), respectively, as
\bea
\label{eq:2nd-velocity-eq-1}
\frac1a \partial_t \left( a \partial^2 \PhiTwo \right) - \frac{3}{2} H^2 \left( a \partial\cdot\vTwo \right) & = & \dot\gamma^{ij} \partial_i \partial_j \PhiOne + H \partial^2 \delta\PhiTwo, \\
\label{eq:2nd-velocity-eq-2}
\partial_t \left( a \partial\cdot\vTwo \right) - \partial^2 \PhiTwo & = & - \partial^2 \delta\PhiTwo.
\eea
Eliminating $\partial\cdot\vTwo$ and re-arranging the equation, we find
\bea
\label{eq:2nd-Phi-massage}
\ddot \Phi^{(2)}  + 4 H \dot \Phi^{(2)} & = & \frac{1}{a^2} \partial^{-2} \left[ \left( \partial^2 \gamma_{ij} \right) \left( \partial^i \partial^j \PhiOne \right) \right] + H \delta\dot\Phi^{(2)}.
\eea
This is to be compared with Eq.~(\ref{eq:2nd-Phi-trace}) to give
$\partial^2\delta\PhiTwo = 3 \partial^{-2} \left[ \left(
\partial^2 \gamma^{ij} \right) \left( \partial_i \partial_j
\PhiOne \right) \right]$. The second source term in
Eq.~(\ref{eq:2nd-Phi-trace}) is suppressed by a factor of
$(K/k)^2$, which is chosen to be small, relative to the first
term.  Therefore, we ignore this term afterwards.

Once $\PhiTwo$ is solved, $\deltaTwo$ and $\vTwo_i$ can be obtained from algebraic relations, 
\bea
\label{eq:2nd-overdensity-algebraic}
\deltaTwo & = & - \frac{2}{3a^2H^2} \partial^2 \PhiTwo + \frac{2}{H} \dot\Phi^{(2)} + 2 \PhiTwo - 2 \delta\PhiTwo + \frac{2}{3a^2H^2} \gamma^{ij} \partial_i \partial_j \PhiOne, \\
\label{eq:2nd-velocity-algebraic}
\frac1a \partial\cdot\vTwo & = & \frac{2}{3a^2H^2} \partial^2 \dot\Phi^{(2)} + \frac{2}{3a^2H^2} H \partial^2 \PhiTwo - \frac{2}{3a^2H^2} H \partial^2 \delta\PhiTwo - \frac{2}{3a^2H^2} \dot\gamma^{ij}\partial_i\partial_j \PhiOne,
\eea
following Eq.~(\ref{eq:2nd-Einstein-time-time}) and
Eq.~(\ref{eq:2nd-Einstein-time-space}). When solving for
$\deltaTwo$, we also ignore the term $2\dot\Phi^{(2)}/H$ in
Eq.~(\ref{eq:2nd-overdensity-algebraic}); it does not grow
$\propto a$ over time, and hence becomes negligible at late
times when $k \gg aH$.

Eq.~(\ref{eq:2nd-Phi-equation}) can be solved by the Green's function. Two independent solutions to the homogeneous part of $\ddot\Phi^{(2)}+4H\dot\Phi^{(2)}=S_{\Phi}$ (where $S_{\Phi}(t)$ denotes the source term) includes a constant solution $\phi_1=1$ and a decaying one $\phi_2=(aH)^5$. The retarded Green's function,
\bea
G_{\mathrm{ret}}\left(t-t'\right) & = & \frac{\phi_1(t) \phi_2(t') - \phi_2(t) \phi_1(t')}{\dot\phi_1(t') \phi_2(t') - \dot\phi_2(t') \phi_1(t')} \Theta\left( t - t' \right) = \frac{2}{5H(t')} \left[ 1 - \left( \frac{a(t)H(t)}{a(t')H(t')} \right)^5 \right] \Theta\left( t - t' \right),
\eea
is then constructed from those two solutions.
With null initial condition imposed, the solution for $\PhiTwo$ is
\bea
\PhiTwo(t) & = & \int^t_0 dt' S_{\Phi}(t') G_{\mathrm{ret}}\left(t-t'\right).
\eea

\section{Alternative derivation using Lagrangian coordinates}
\label{app:lagrangian-coordinates}

In this Appendix we provide an alternative derivation, using
Lagrangian coordinates for a collection of freely-falling
particles, of Eq.~(\ref{eq:delta-2nd-order}).

The matter distribution can be visualized as a collection of
a huge number of non-relativistic matter particles of equal masses
that fill the space.  For collisionless matter, one
can then track the position of individual particle---i.e., the
Lagrangian coordinate---along the geodesic. 

Consider a particle at comoving position $x^i$ at early times
($t \rightarrow 0$).  Let $s^i(x^j,t)$ be the comoving
displacement of that particle with respect to its initial
position at any later time $t$. The matter overdensity arises
because the Lagrangian volume element differs from point to
point, and we have
\bea
\delta & = & \left( 1+\delta_p \right) \det\left[ \frac{\partial s^i}{\partial x^j} \right]^{-1} - 1 \approx \delta_p - \partial\cdot s,
\eea
where $\delta_p$ is the overdensity at initial time. Note that
keeping terms linear in $s^i$ suffices; since $\gamma_{ij}$ does
not deflect comoving massive particles, $s^i \sim
\mathcal{O}(\Phi)$, and terms higher-order in $s^i$ are at least
$\mathcal{O}(\Phi^2)$, which are consistently ignored throughout
this paper.

The 4-displacement $s^{\mu}\equiv(t,s^i)$ can be solved from the
geodesic equation,
\bea
\label{eq:lagrangian-geodesic}
\frac{d^2 s^{\mu}}{d \lambda^2} & = & - \Gamma^{\mu}_{\alpha\beta} \frac{ds^{\alpha}}{d\lambda} \frac{ds^{\beta}}{d\lambda},
\eea
where $\lambda$ is the proper time.
The $\Gamma^{\mu}_{\alpha\beta}$'s are calculated from
Eq.~(\ref{eq:perturbed-metric}). The geodesic equation must be
supplemented by the equation describing how the matter distribution
generates the gravitational potential $\Phi$. On subhorizon
scales $k \gg aH$, the Poisson equation,
\bea
\label{eq:lagrangian-poisson}
- \frac{1}{a^2} \left( \delta^{ij} - \gamma^{ij} \right) \partial_i \partial_j \Phi & = & 4\pi G \rho_m \delta = 4\pi G \rho_m \left( \delta_p - \partial\cdot s \right),
\eea
does the job.  The tensor perturbation $\gamma_{ij}$ enters the
left-hand side because the Poisson equation holds only in a ``locally
Newtonian'' frame where $\gamma_{ij}$ is not felt. This is also
validated by the subhorizon limit of
Eq.~(\ref{eq:Einstein-time-time-perturb}). Furthermore, in the
Newtonian limit we always assume $\Psi=-\Phi$.

Given the evolution of $\gamma_{ij}$ in
Eq.~(\ref{eq:tensor-evolve-linear}),
Eqs.~(\ref{eq:lagrangian-geodesic}) and
(\ref{eq:lagrangian-poisson}) can be solved
perturbatively. Following the spirit of
Sec.~\ref{sec:post-inflation}, we keep the terms $s^i$ and
$\gamma_{ij}$ linear in $\Phi$, as well as terms of
$\mathcal{O}(\Phi\gamma,s\gamma)$, but not terms quadratic in
$\gamma_{ij}$ {\it or} in potential/displacement.  We split
solutions into a linear part in the absence of $\gamma_{ij}$,
and a nonlinear correction of $\mathcal{O}(s\gamma)$.  In
analogy to the notation of Sec.~\ref{sec:post-inflation}, we
write $s^i=s^{(1)i}+s^{(2)i}$ and $t=t^{(1)}+t^{(2)}$. 

\subsection{Linear solutions in the absence of tensor}
\label{subapp:lagrangian-coordinates-linear}

The linear solutions can be obtained by taking
$\gamma_{ij}=0$. In order to solve for $s^{(1)i}$, setting
$t=\lambda$ suffices. The geodesic equation gives
\bea
\ddot s^{(1)i} & = & - \frac{1}{a^2} \partial^i \Psi.
\eea
Taking the divergence, and supplemented with the Poisson
equation at linear order, we have
\bea
\frac{d}{dt} \left[ a^2 \frac{d}{dt} \left( \partial \cdot s^{(1)} \right) \right] & = & \partial^2 \PhiOne, \\
\label{eq:lagrangian-poisson-linear}
- \partial^2 \PhiOne & = & - 4\pi G a^2 \rho_m \partial\cdot s^{(1)},
\eea
which combine to give
\bea
\label{eq:lagrangian-displacement-linear-2ndOPE}
\frac{d}{dt} \left[ a^3 \frac{d}{dt} \left( \frac{ \partial \cdot s^{(1)} }{a} \right) \right] + a^3 H \frac{d}{dt}  \left( \frac{ \partial \cdot s^{(1)} }{a} \right) & = & 0.
\eea
Ignoring the decaying solution, we find $(\partial \cdot
s^{(1)})/a=$constant, and hence $\PhiOne\equiv \Phi_p$ is constant
over time. From Eq.~(\ref{eq:lagrangian-poisson-linear}) and
the time component of Eq.~(\ref{eq:lagrangian-geodesic}), we
then find
\bea
s^{(1)i} = \frac{2}{3a^2H^2} \partial^i \Phi_p,\qquad t^{(1)} = 0.
\eea

\subsection{Nonlinear corrections due to tensor}
\label{subapp:lagrangian-coordinates-2nd}

Since $t^{(1)}=0$, it turns out that when solving for $s^{(2)i}$
we can still identify $t=\lambda$. Then we can take the
$\mathcal{O}(\Phi\gamma,s\gamma)$ part of the spatial geodesic
equation,
\bea
\frac{d^2 s^i} {dt^2} & = & - \Gamma^i_{00} - 2 \Gamma^i_{j0} \frac{ds^j}{dt} - \Gamma^i_{jk} \frac{ds^j}{dt} \frac{ds^k}{dt}.
\eea
to obtain
\bea
\ddot s^{(2)i} & = & \frac{1}{a^2} \partial^i \PhiTwo - \frac{1}{a} \left( \dot w^i + H w^i \right) - \frac{1}{a^2} \gamma^{ij} \partial_j \PhiOne - 2 H \dot s^{(2)i} - \dot\gamma^i_j \dot s^{(1)j},
\eea
where we have used $\dot\Phi^{(1)}=0$ to simplify. We then take
the divergence, and use $\partial\cdot w=0$ and $\partial_i \gamma^{ij}=0$ to obtain
\bea
\label{eq:lagrangian-geodesic-2nd}
\frac{d}{dt} \left[ a^2 \frac{d}{dt} \left( \partial \cdot s^{(2)} \right) \right] & = & \partial^2 \PhiTwo - \gamma^{ij} \partial_i \partial_j \PhiOne - \frac{2}{3H} \dot\gamma^{ij} \partial_i \partial_j \PhiOne.
\eea
The $\mathcal{O}(\Phi\gamma,s\gamma)$ part of the Poisson equation is
\bea
\label{eq:lagrangian-poisson-2nd}
- \frac{1}{a^2} \partial^2 \PhiTwo + \frac{1}{a^2} \gamma^{ij} \partial_i \partial_j \PhiOne & = & - 4\pi G \rho_m \partial \cdot s^{(2)}.
\eea
We then combine Eq.~(\ref{eq:lagrangian-geodesic-2nd}) and
Eq.~(\ref{eq:lagrangian-poisson-2nd}) to eliminate $\PhiTwo$,
and obtain
\bea
\frac{d}{dt} \left[ a^2 \frac{d}{dt} \left( \partial \cdot s^{(2)} \right) \right] - \frac{3a^2H^2}{2} \partial \cdot s^{(2)}   & = & - \frac{2}{3H} \dot\gamma^{ij} \partial_i \partial_j \PhiOne,
\eea
or in terms of conformal time,
\bea
\label{eq:lagrangian-displacement-2nd}
\frac{d^2 }{d\eta^2} \left( \partial\cdot s^{(2)} \right) 
+ \frac{2}{\eta}\frac{d}{d\eta} \left( \partial\cdot s^{(2)} \right) - \frac{6}{\eta^2} \left( \partial\cdot s^{(2)} \right) & = & - \frac{2}{3H} \dot\gamma^{ij} \partial_i \partial_j \Phi_p.
\eea
The most general solution can be written as
\bea
\partial\cdot s^{(2)} & = & \left( \partial\cdot s^{(2)} \right)_{\rm homo} + \left( \partial\cdot s^{(2)} \right)_{\rm spec},
\eea
where $\left( \partial\cdot s^{(2)} \right)_{\rm homo}$ solves the
homogeneous part of Eq.~(\ref{eq:lagrangian-displacement-2nd}),
while $\left( \partial\cdot s^{(2)} \right)_{\rm spec}$ is a special
solution that solves the full equation but vanishes at $\eta=0$. 

The homogeneous part of
Eq.~(\ref{eq:lagrangian-displacement-2nd}) has two independent
solutions, $\phi_1(\eta)=\eta^2$ and
$\phi_2(\eta)=1/\eta^3$. Obviously, $\left( \partial\cdot
s^{(2)} \right)_{\rm homo}$ has to be $\propto \eta^2$. In fact, it
has to be the unique solution for the case of infinite tensor
wavelength, $K\rightarrow 0$ and $\gamma_{ij}\equiv
\gamma_{p,ij}$. In that case, the same derivation to conclude
$(\partial \cdot s^{(1)})/a=$constant from
Eq.~(\ref{eq:lagrangian-displacement-linear-2ndOPE}) leads to
$(\partial \cdot s^{(2)})/a=$constant, and hence
$\dot\Phi^{(2)}=0$. From the initial condition
$\Phi(t=0)=\Phi_p$, it must be that $\PhiTwo=0$ for infinite
tensor wavelength, and hence,
\bea
\left(  \partial\cdot s^{(2)} \right)_{\rm homo} & = & \frac{2}{3a^2H^2} \gamma^{ij}_p \partial_i \partial_j \Phi_p.
\eea

To find $\left( \partial\cdot s^{(2)} \right)_{\rm spec}$ for finite
tensor wavelength, we construct the retarded Green's function,
\bea
G_{\mathrm{ret}}\left(\eta-\eta'\right) & = & \frac{\phi_1(\eta) \phi_2(\eta') - \phi_2(\eta) \phi_1(\eta')}{\dot\phi_1(\eta') \phi_2(\eta') - \dot\phi_2(\eta') \phi_1(\eta')} \Theta\left( \eta - \eta' \right) = \frac{\eta^2}{5\eta'} \left[ 1 - \left(\frac{\eta'}{\eta}\right)^5 \right] \Theta\left( \eta - \eta' \right),
\eea
and then find
\bea
\left(  \partial\cdot s^{(2)} \right)_{\rm spec} & = & \int^{\eta}_0 d\eta' \left( - \frac{2}{3H(\eta')} \dot\gamma^{ij}(\eta') \partial_i \partial_j \Phi_p \right) G_{\mathrm{ret}}\left(\eta-\eta'\right) \nn\\
& = & - \frac{2}{3a^2H^2} \mathcal{S}_N(K) \gamma^{ij}_p \partial_i \partial_j \Phi_p,
\eea
where we define the function
\bea
\label{eq:kernel-SN-def}
\mathcal{S}_N(K) & \equiv & \int^{K\eta}_0 d(K\eta') \frac{2}{5} \frac{ \partial \mathcal{T}_{\gamma}(K\eta')}{\partial (K\eta')} \left[ 1 - \left(\frac{K\eta'}{K\eta} \right)^5 \right].
\eea
Combining all the results, we are able to write (in Fourier space)
\bea
\label{eq:lagrangian-overdensity-2nd-result}
\delta = \delta_p - \partial\cdot s^{(1)} - \partial\cdot s^{(2)} = \delta_p + 2 \mathcal{T}_{\delta}(k) \Phi_p \left[ 1 - \frac12 \frac{d\ln \mathcal{T}_{\delta}}{d\ln k} \gamma_{p,ij} \hat k^i \hat k^j - \mathcal{S}_N(K) \gamma_{p,ij} \hat k^i \hat k^j  \right],
\eea
given the linear-extrapolation factor in the subhorizon limit
$\mathcal{T}_{\delta}(k)=2k^2/(3a^2H^2)$. The right hand side
can be simply evaluated at the initial location $x^i$ even if the
test particle has moved by $s^i$; the difference made to
$\delta$ is $\sim \Phi\,s \sim \mathcal{O}(\Phi^2)$ and hence
can be ignored.
For the subhorizon density modes that we are considering here, 
$\delta_p =2 \Phi_p \ll 2\mathcal{T}_\delta(k)\Phi_p$, and 
Eq.~(\ref{eq:lagrangian-overdensity-2nd-result}) exactly agrees
with Eq.~(\ref{eq:delta-2nd-order}), given 
a correspondence between $S_N(K)$ here and $S(K)$ in Eq.~(\ref{eq:SK}):
\bea
\mathcal{S}_N(K) + \frac12 \frac{d\ln\mathcal{T}_{\delta}}{d\ln k} \left( 1 - \mathcal{T}_{\gamma}(K) \right) = \mathcal{S}(K),
\eea
where $d\ln\mathcal{T}_{\delta}/d\ln k=2$ in the subhorizon limit $k\gg aH$. 

With $S_N(K)$ we have calculated in this Section, we can rewrite the fractional
density perturbation in Eq.~(\ref{eq:delta-2nd-order}) as 
\be
\delta(k) 
= 2\mathcal{T}_\delta(k)
\left(
1- \frac12 \frac{d\ln T_\delta(k)}{d\ln k} \gamma_p^{ij}\hat{k}_i\hat{k}_j
\right) \Phi_p
-
2\mathcal{T}_\delta(k)\gamma_p^{ij}\hat{k}_i\hat{k}_j \Phi_p S_N(K),
\ee
and the corresponding matter power spectrum in the local coordinate
Eq.~(\ref{eq:tilde_Pdelta_kxc}) as 
\begin{align}
\tilde{P}_\delta(\veck)
=& 4\mathcal{T}_\delta^2(k)
\left[
P_{\Phi}(k)
- 
\left(\frac12 \frac{d\ln P_{\delta}(k)}{d\ln k}
+
2S_N(k)
\right)
\gamma_p^{ij}\hat{k}_i\hat{k}_jP_{\Phi}(k)
\right].
\end{align}
Finally, these changes are translated to the quadrupole 
in the observed power spectrum Eq. (\ref{eq:quadrupole-tensor}) as
\bea
     && \mathcal{Q}_{ij} = 
     - \left(\frac12 \frac{d\ln P_{\delta}}{d\ln k}  
            + 2 \mathcal{S}_N(K) \right) \gamma_{p,ij} 
     - \frac{d\ln P_{\delta}}{d\ln k} 
     \left(
     \partial_{(i} \Delta x_{j)} - \frac13 \delta_{ij}
     \partial\cdot\Delta x \right).
\eea
We further re-arrange the quadrupole moments $\mathcal{Q}_{ij}$ as 
\bea
\label{eq:FNC-suggestive-form}
     && \mathcal{Q}_{ij} = 
     - \frac12 \frac{d\ln P_{\delta}}{d\ln k} 
       \left( 1 - \mathcal{T}_{\gamma}\right) \gamma_{p,ij} 
            + 2 \mathcal{S}_N(K) \gamma_{p,ij} 
     - \frac{d\ln P_{\delta}}{d\ln k} 
     \left(
		\frac12 \gamma_{ij} + 
     \partial_{(i} \Delta x_{j)} - \frac13 \delta_{ij}
     \partial\cdot\Delta x \right).
\eea
The third term in the parenthesis takes into account the projection effect,
as it is the observed shear component including
the metric shear \cite{Dodelson:2003bv,Schmidt:2012nw,Schmidt:2012ne}
that arises from the transformation between 
local and global coordinates.
That is, the third term is the quadrupole observed from a galaxy power spectrum
which is isotropic in the local frame.
Therefore, the first two terms should be interpreted as the ``tidal effect'' 
from the long-wavelength tensor mode to the locally observed 
matter power spectrum at the time when observed galaxies emitted 
photons.

\section{Galaxy overdensity in observed coordinates}
\label{app:galaxy-observed}

Since the number of galaxies is conserved in whatever coordinates one uses, we can relate number density measured in the comoving FRW coordinates $n_g$ to that measured in observed coordinates $\tilde n_g$ through
\bea
\label{eq:conservation-no-galaxy}
a^3\left(\tilde t\right) \tilde n_g\left( \tilde\bfx,\tilde t \right) d^3 \tilde x = a^3(t) \left[\det\left(\delta_{ij} + \gamma_{ij}\right)\right]^{1/2} n_g( \bfx,t ) d^3 x,
\eea
where the Jacobian determinant is unity for traceless $\gamma_{ij}$. We then define galaxy overdensities in both coordinates w.r.t. the average number density $\bar n_g$ as expected from the homogeneous background cosmology,
\bea
n_g\left( \bfx, t \right) & = & \bar n_g(t) \left( 1 + \delta_g \left( \bfx, t \right) \right), \\
\tilde n_g\left( \tilde\bfx, \tilde t \right) & = & \bar n_g(\tilde t) \left( 1 + \tilde\delta_g \left( \tilde\bfx, \tilde t \right) \right).
\eea
We can calculate $(a^3(t) \bar n_g(t))/(a^3\left(\tilde t\right)
\bar n_g\left(\tilde t\right)) = 1 + b_e H \Delta t$, where $b_e
\equiv ( d\ln a^3 n_g )/(d \ln a)$ can be measured for a
specific galaxy sample. Besides, we have $d^3 x/d^3 \tilde x = 1
+ \partial_i \Delta x^i$. Combining all the pieces, we find
\bea
\tilde \delta_g - \delta_g & = & \left( \Delta_t \partial_t + \Delta x^i \partial_i \right) \delta_g + \left( b_e H \Delta t + \partial_i \Delta x^i \right) +  \left( b_e H \Delta t + \partial_i \Delta x^i \right) \delta_g,
\eea
which is Eq.~(\ref{eq:galaxy-overdensity-transform}).

\section{Local galaxy power spectrum in observed coordinates}
\label{app:galaxy-observed-power-spectrum}

The galaxy power spectrum on small scales can be written as
\bea
\label{eq:observed-power-spectrum}
\tilde P_g(\bfk) & = & \int d^3\delta\bfx e^{-i\bfk\cdot\delta\bfx} \VEV{\tilde\delta_g\left(\bfx_1\right) \tilde\delta_g\left(\bfx_2\right)}_{\gamma}.
\eea
Inside the integral, we correlate two points $\bfx_1$ and $\bfx_2$ separated by $\delta\bfx$ with midpoint $\bfx_c$, so that $\bfx_{1,2}=\bfx_c\mp\delta\bfx/2$. Assume that $|\delta\bfx|$ is small compared to both $|\bfx_c|$ (the distance from the observer to the galaxy) and the variation scale of tensor perturbations $1/K$.  

Next, we apply Eq.~(\ref{eq:galaxy-overdensity-observed}), together with
\bea
\frac{\partial}{\partial \bfx_1} = \frac12 \frac{\partial}{\partial \bfx_c} - \frac{\partial}{\partial \delta\bfx}, \qquad \frac{\partial}{\partial \bfx_2} = \frac12 \frac{\partial}{\partial \bfx_c} + \frac{\partial}{\partial \delta\bfx}.
\eea
In particular, for the $\partial_i\delta$ term in
Eq.~(\ref{eq:galaxy-overdensity-observed}), we expand to
linear order in $\delta\bfx$, 
\bea
\Delta x^i(\bfx_1) \frac{\partial}{\partial x^i_1} \delta(\bfx_1) & = & \left( \Delta x^i - \frac12 \delta x^j \partial_j \Delta x^i \right) \left( \frac12 \frac{\partial}{\partial x_c^i} - \frac{\partial}{\partial \delta x^i} \right) \delta(\bfx_1), \nn\\
\Delta x^i(\bfx_2) \frac{\partial}{\partial x^i_2} \delta(\bfx_2) & = & \left( \Delta x^i + \frac12 \delta x^j \partial_j \Delta x^i \right) \left( \frac12 \frac{\partial}{\partial x_c^i} + \frac{\partial}{\partial \delta x^i} \right) \delta(\bfx_2),
\eea
as needed to find a
quadrupole in the local power spectrum\footnote{The higher-order terms
in the expansion, are negligibly small and give higher-order
multipoles.}
where $\Delta x^i$ and $\partial_j\Delta x^i$ are to be computed at the midpoint $\bfx_c$. These can be recast in Fourier space, where we trade the separation $\delta\bfx$ (not the midpoint $\bfx_c$) for the conjugated wavevector $\bfk$,
\bea
\Delta x^i(\bfx_1) \frac{\partial}{\partial x^i_1} \delta(\bfx_1) & = & \int d^3\bfk e^{i\bfk\cdot(-\delta\bfx/2)} \left( \Delta x^i + i \partial_j \Delta x^i \frac{\partial}{\partial k_j} \right) \left( \frac12 \frac{\partial}{\partial x_c^i} + \frac i2 k_i \right) \delta(\bfk), \nn\\
\label{eq:derivative-term-expansion-Fourier}
\Delta x^i(\bfx_2) \frac{\partial}{\partial x^i_2} \delta(\bfx_2) & = & \int d^3\bfk e^{i\bfk\cdot(\delta\bfx/2)} \left( \Delta x^i + i \partial_j \Delta x^i \frac{\partial}{\partial k_j} \right) \left( \frac12 \frac{\partial}{\partial x_c^i} + \frac i2 k_i \right) \delta(\bfk),
\eea
On the other hand, the $\left( b_e H \Delta t + \partial_i \Delta x^i \right)$ term can be just evaluated at the midpoint $\bfx$, since the correction starts only at quadratic order in $\delta\bfx$.

Whenever a term is explicitly multiplied by a quantity first-order in $\gamma_{ij}$, we can plug in the zeroth-order isotropic matter power spectrum $\VEV{\delta(\bfk)\delta(\bfk')}_0 = (2\pi)^3 \delta^{(3)}(\bfk+\bfk') P_{\delta}(k)$, which does not depend on $\bfx_c$ from statistical homogeneity in the absence of tensor perturbations. Then we are able to combine Eq.~(\ref{eq:galaxy-overdensity-observed}) and Eq.~(\ref{eq:derivative-term-expansion-Fourier}) to derive
\bea
\tilde P_g(\bfk;\bfx_c) & = & b^2_g \left[ P_{\delta}(\bfk;\bfx_c) - \left(\partial_j \Delta x^i\right) \frac{\partial}{\partial k_j} k_i P_{\delta}(k) + 2\left( b_e H \Delta t + \partial_i \Delta x^i \right) P_{\delta}(k)  \right] \nn\\
& = & b^2_g \left[ P_{\delta}(\bfk;\bfx_c) - \left(\partial_j \Delta x^i\right) k_i \frac{\partial}{\partial k_j} P_{\delta}(k) + \left( 2 b_e H \Delta t + \partial_i \Delta x^i \right) P_{\delta}(k)  \right] \nn\\
& = & b^2_g \left[ P_{\delta}(\bfk;\bfx_c) - \left(\partial_j \Delta x_i\right) \hat k^i \hat k^j \frac{d\ln P_{\delta}(k)}{d\ln k} P_{\delta}(k) + \left( 2 b_e H \Delta t + \partial_i \Delta x^i \right) P_{\delta}(k)  \right].
\eea
This gives Eq.~(\ref{eq:local-galaxy-power-spectrum-result}). It is understood that background quantities (e.g. $a$ and $H$) and linear-order quantities in $\gamma_{ij}$ (e.g. $\Delta x^i$ and $\Delta t$) are computed at the midpoint $\bfx_c$, whose values are representative across the local volume.

\section{Calculation of the galaxy power quadrupole from long-wavelength tensor perturbations}
\label{app:galaxy-power-quadrupole-calc}

Following the formalism of Ref.~\cite{Dai:2012bc}, we expand the
tensor perturbation field (at primordial time) in terms of
total-angular-momentum (TAM) waves,
\bea
\gamma_{p,ij}(\bfx) & = & \int \frac{K^2dK}{(2\pi)^3} \sum_{J=2}^{\infty} \sum_{M=-J}^J \sum_{\alpha=TE,TB} \gamma^{\alpha}_{p,JM}(K) (4\pi i^J) \Psi^{\alpha,K}_{(JM)ij}(\bfx).
\eea
From the Fourier-space power spectrum, Eq.~(\ref{eq:tensor-power-spectrum-primordial}), the TAM coefficients will satisfy
\bea
\VEV{\gamma^{\alpha}_{p,JM}(K) \gamma^{\alpha'*}_{p,J'M'}(K')} = \frac{(2\pi)^3}{K^2} \delta_D(K-K') P_{\gamma}(K) \delta_{JJ'} \delta_{MM'} \delta_{\alpha \alpha'}.
\eea
The quadrupole tensor measured at $\bfx$ from a single TAM mode of wavenumber $K$ and total-angular-momentum $J$ and $M$, analogous to Eq.~(\ref{eq:quadrupole-tensor}), can be expanded in terms of tensor spherical harmonics,
\bea
\mathcal{Q}_{ij}(\bfx_c) & = & \gamma^{TE}_{p,JM}(K) \sum_{\alpha=L,VE,TE} \mathcal{Q}^{\alpha}_J(K,z) Y^{\alpha}_{(JM)ij}(\bfn) + \gamma^{TB}_{p,JM}(K,z) \sum_{\alpha=VB,TB} \mathcal{Q}^{\alpha}_J(K) Y^{\alpha}_{(JM)ij}(\bfn),
\eea
where the direction $\bfn$ on the sky is exactly the direction
of $\bfx_c$ as measured from the observer, and the redshift $z$
corresponds to the comoving distance $r=|\bfx|$ assuming a
background cosmology.  Then in
Eq.~(\ref{eq:expression-quadrupole-rms}), $\mathcal{Q}^2_J(K,z)
= \sum_{\alpha} \left| \mathcal{Q}^{\alpha}_J(K,z) \right|^2$,
where we sum over the five types $\alpha=L,VE,VB,TE,TB$ of
tensor spherical harmonics.

Below we present the explicit expressions for $\mathcal{Q}^{\alpha}_J(K)$. First, we define coefficients,
\bea
\kappa_1 & = & - \frac12 \left( \frac{d\ln P_{\Phi}}{d\ln k} + \mathcal{T}_{\gamma}(K) \frac{d\ln \mathcal{T}^2_{\delta}}{d\ln k} \right), \\
\kappa_2 & = & -  2 \mathcal{S}(K), \\
\kappa_3 & = & - \left( \frac{d\ln P_{\Phi}}{d\ln k} + \frac{d\ln \mathcal{T}^2_{\delta}}{d\ln k} \right),
\eea
which correspond to primordial, nonlinear mode-coupling, and
projection effects, respectively. We then have
\bea
\mathcal{Q}^L_J(K,z) & = &  - \left( \kappa_1 + \kappa_2 \right) j^{(L,TE)}_{J,t}(Kr) + \kappa_3 \left( \sqrt{\frac 23} \left( \frac 1r - \partial_r + a \partial_t \right) \Delta x^{TE}_{\parallel} + \sqrt{\frac{J(J+1)}{6}} \frac{\Delta x^{TE}_{\perp}}{r} \right), \\
\mathcal{Q}^{VE}_J(K,z) & = &  - \left( \kappa_1 + \kappa_2 \right) j^{(VE,TE)}_{J,t}(Kr) + \kappa_3 \left( - \frac{1}{\sqrt 2} \left( \frac 1r - \partial_r + a \partial_t \right) \Delta x^{TE}_{\perp} - \sqrt{\frac{J(J+1)}{2}} \frac{\Delta x^{TE}_{\parallel}}{r} \right), \\
\mathcal{Q}^{TE}_J(K,z) & = &  - \left( \kappa_1 + \kappa_2 \right) j^{(TE,TE)}_{J,t}(Kr) + \kappa_3 \sqrt{\frac{(J-1)(J+2)}{2}} \frac{\Delta x^{TE}_{\perp}}{r},
\eea
from $TE$-type TAM modes, and
\bea
\mathcal{Q}^{VB}_J(K) & = &  - i \left( \kappa_1 + \kappa_2 \right) j^{(VB,TB)}_{J,t}(Kr) + i \kappa_3 \left( - \frac{1}{\sqrt 2} \left[ \left( \frac 1r - \partial_r + a \partial_t \right) \Delta x^{TB}_{\perp} \right] \right), \\
\mathcal{Q}^{TB}_J(K) & = &  - i \left( \kappa_1 + \kappa_2 \right) j^{(TB,TB)}_{J,t}(Kr) + i \kappa_3 \sqrt{\frac{(J-1)(J+2)}{2}} \frac{\Delta x^{TB}_{\perp}}{r},
\eea
from $TB$-type TAM modes. Here $r$ is the comoving distance to the source, and the radial functions $j^{(\alpha,\alpha')}_{J,t}(x)$ are given in Eqs.~(22), (24), and (25) of Ref.\cite{Dai:2012ma}. Finally, terms involving components of $\Delta x^i$, which can be calculated from Eqs.~(\ref{eq:shift-parallel}) and (\ref{eq:shift-perp}), are given by
\bea
\frac{\Delta x^{TB}_{\perp}}{r} & = & \sqrt{\frac{(J-1)(J+2)}{2}} \mathcal{I}_1, \\
\left( \frac1r - \partial_r + a \partial_t \right) \Delta x^{TB}_{\perp} & = & - \sqrt{\frac{(J-1)(J+2)}{2}} \mathcal{T}_{\gamma} \frac{j_J(Kr)}{Kr}, \\
\frac{\Delta x^{TE}_{\parallel}}{r} & = & \sqrt{\frac{(J+2)!}{2(J-2)!}} \left( -\frac12 \frac{1}{Kr} \mathcal{I}_1 - \frac{1}{2aHr} \mathcal{I}_2 \right), \\
\left( \frac1r - \partial_r + a \partial_t \right) \Delta x^{TE}_{\parallel} & = & \sqrt{\frac{(J+2)!}{2(J-2)!}} \frac12 \left[ \left( \mathcal{T}_{\gamma} + \frac{K}{aH} \frac{\partial \mathcal{T}_{\gamma}}{\partial (K\eta)}  \right) \frac{j_J(Kr)}{(Kr)^2} - \left( \left( 1+\frac{\dot H}{H^2} \right) + \frac{1}{aHr} \right) \mathcal{I}_2 - \frac{\mathcal{I}_1}{Kr} \right], \\
\frac{\Delta x^{TE}_{\perp}}{r} & = &  \sqrt{\frac{(J-1)(J+2)}{2}} \left( - \frac{\mathcal{T}_{\gamma,o}}{10} \delta_{J,2} + \mathcal{I}_3 \right), \\
\left( \frac1r - \partial_r + a \partial_t \right) \Delta x^{TE}_{\perp} & = &  \sqrt{\frac{(J-1)(J+2)}{2}} \left[ - \frac{\mathcal{T}_{\gamma}}{Kr} \left( j_J'(Kr) + \frac{j_J(Kr)}{Kr} \right) + \frac{J(J+1)}{2} \frac{\mathcal{I}_1}{Kr} \right],
\eea
where we define line-of-sight integrals,
\bea
\mathcal{I}_1 & \equiv & \int^{Kr}_0 dx \mathcal{T}_{\gamma} \frac{j_J(x)}{x^2}, \\
\mathcal{I}_2 & \equiv & \int^{Kr}_0 dx \frac{\partial \mathcal{T}_{\gamma}}{\partial (K \eta)} \frac{j_J(x)}{x^2}, \\
\mathcal{I}_3 & \equiv & \int^{Kr}_0 dx \frac{\mathcal{T}_{\gamma}}{x^2} \left( j_J'(x) + \frac{j_J(x)}{x} - \frac{J(J+1)}{2} \left( 1 - \frac{x}{Kr} \right) \frac{j_J(x)}{x} \right).
\eea

\end{widetext}


\end{document}